# The Dawn of Dust Astronomy

Eberhard Grün[1,2*], Harald Krüger[3], and Ralf Srama[4]

[1] *Max-Planck-Institut für Kernphysik, Heidelberg, Germany*

[2] *LASP, University of Colorado, Boulder, CO, USA*

[3] *Max-Planck-Institut für Sonnensystemforschung, Göttingen, Germany*

[4] *Institut für Raumfahrtsysteme Universität Stuttgart, Germany*

[*] Corresponding author. Email: harald.krueger@mps.mpg.de, Tel.: +49-551-384979-234.

## Abstract

We review the development of dust science from the first ground-based astronomical observations of dust in space to compositional analysis of individual dust particles and their source objects. A multitude of observational techniques is available for the scientific study of space dust: from meteors and interplanetary dust particles collected in the upper atmosphere to dust analyzed in situ or returned to Earth. In situ dust detectors have been developed from simple dust impact detectors determining the dust hazard in Earth orbit to dust telescopes capable of providing compositional analysis and accurate trajectory determination of individual dust particles in space. The concept of Dust Astronomy has been developed, recognizing that dust particles, like photons, carry information from remote sites in space and time. From knowledge of the dust particles' birthplace and their bulk properties, we learn about the remote environment out of which the particles were formed. Dust Observatory missions like Cassini, Stardust, and Rosetta study Saturn’s satellites and rings and the dust environments of comet Wild 2 and comet Churyumov-Gerasimenko, respectively. Supplemented by simulations of dusty processes in the laboratory we are beginning to understand the dusty environments in space.

# 1. Introduction

Dusty phenomena in space have been studied by astronomers for centuries: the zodiacal light, comets, and meteors. The zodiacal light is a prominent phenomenon observable by the human eye in the morning and evening sky in non-polluted areas on Earth. Already in 1683 the Italian astronomer Giovanni Domenico Cassini presented the correct explanation: it is sunlight scattered by dust particles orbiting the Sun. Comets shed large amounts of dust during their passages through the inner solar system that is mostly visible in the cometary coma and dust tail. The third dusty phenomenon visible with the naked eye, meteors, can be observed during any clear night, but there are special periods, so called meteor showers or meteor storms, when the rate is thousand times enhanced. It was the coincidence between such meteor showers and the simultaneous apparitions of comets that suggested their genetic relation. Triangulation of meteor trails in the atmosphere, especially during meteor showers, confirmed the relation of some meteor streams to comets. The physical interrelation of the three phenomena was illuminated by the work of the American astronomer Fred Whipple in the early 1950s (Whipple, 1949, 1951, 1955).

Interstellar dust became a topic of astrophysical research in the early 1930s when the existence of extinction, weakening, and scattering of starlight in the interstellar medium (ISM) was realized (cf. Trümpler, 1930). Since then astronomical observations provided information about the properties of the dust in the ISM. Dense clouds of interstellar dust are visible as dark nebulae against the brighter background of the Milky Way. The distance dependent attenuation of blue star light is much stronger than that of red light; therefore, extinction causes distant stars to appear redder than expected, a phenomenon referred to as interstellar reddening. Broad absorption features at infrared (IR) wavelengths are clues to the composition of interstellar dust grains, e.g. the water ice feature at 3.1 μm, and silicate features at 10 and 18 μm. Dark interstellar clouds are frequently related to star formation regions. Here, dust plays an important role by radiative cooling of collapsing clouds and eventually providing seeds for condensation and accretion of larger bodies that are building blocks of planetary objects (Dorschner, 2001). Even from expanding shells of supernova explosions dust became introduced into primitive solar system material (Leitner and Hoppe, 2019) .

The ultimate sources of dust are stars breeding heavy elements by nuclear fusion. Dust particles bear information on their formation process, and any subsequent processes which they became part of. In interstellar space, heavy elements, i.e. elements heavier than helium, often reside in dust grains. Together with their corresponding atomic and ionized species they trace stellar evolution in our galaxy.

In 2016, during the final phases of the Rosetta mission around comet 67P/Churyumov-Gerasimenko, the spacecraft flew sufficiently close to the nucleus so that several instruments simultaneously observed outbursts of gas and dust (Grün et al., 2016, Agarwal et al., 2017, and references therein). Monitoring instruments, like the star sensors STR, the dust impact detector GIADA, and the gas sensor ROSINA/COPS, were the first to detect outbursts at the cometary surface. The STR cameras observed stray light from the emitted dust cloud in their fields-of-view, GIADA detected impacts of individual sub-millimetre sized dust particles on its sensors and determined their mass and speed, and COPS recorded the enhanced gas flux from the outburst but also the gas puffs from sublimating icy dust particles entering the pressure sensor. The Alice UV spectrometer recorded signs of gas and dust emissions from the nucleus. In some cases the OSIRIS cameras could identify the outburst location on the nucleus and observe the expanding dust cloud and streaks of centimetre-sized particles passing near the camera. The atomic force microscope instrument MIDAS provided microscopic and morphologic information on collected submicrometer-sized grains, and the COSIMA dust analyzer obtained structural and compositional information on collected sub-millimetre sized grains (Fig. 1). Also, the plasma instruments RPC recorded an effect of the expanding gas and dust cloud. By the combination of this wealth of information on the solid component of the nucleus, we can obtain a better understanding of this comet and the processes occurring on its surface. This described episode demonstrates the many facets that the study of dust and its interrelation with its environment has. Rosetta is a dust observatory mission that provided one of the most recent highlights in dust astronomy.

Since the beginning of space flight in 1957 the interest in space dust has enormously expanded. The first questions asked were: How much dust is in near-Earth space and how big is it? The reasons were obvious: Travelling with a typical speed of 20 km $s^{-1}$, even sub-millimetre-sized meteoroids pose a serious threat to space vehicles. Most early satellites carried dust detectors in order to characterize the meteoroid hazard for space flight. The long path from these early days of simple in situ dust detection to the present days of dust astronomy is portrayed in this article.

Since the early 1960s the community of astronomers and dust scientists met and exchanged their data and ideas at big international conferences, like COSPAR (Committee on Space Research) and IAU (International Astronomical Union) meetings. Soon it was recognized that it was mutually advantageous to bring together these partially overlapping communities of scientists. Therefore, specialized international dust meetings regularly took place at varying locations which produced significant proceedings volumes. In later years dust science spread to broader topics and was discussed at meetings of the Meteoritical Society, at geophysical (AGU, EGS) and planetary science meetings (DPS, EPSC, ACM).

In 1978 Tony McDonnell took up the task to assemble a comprehensive review of the "Cosmic Dust" field. The book (McDonnell, 1978) covered a wide range of topics: from interstellar dust to comets, meteors, and zodiacal light, over lunar craters to dust particles collected in the atmosphere, from dust dynamics to laboratory simulation and impact physics. About 20 years later, in 2001, a new version of a comprehensive review of the field was assembled by Eberhard Grün, Bo Gustafson, Stanley Dermott and Hugo Fechtig (Grün et al., 2001). Besides the classical topics of cometary, interplanetary and interstellar dust, instrumentation and dust dynamics, it covered new subjects like near-Earth dust and space debris, dusty rings, thermal properties of dust and physical processes on interplanetary dust, and a review of comprehensive models of interplanetary dust. The book also contained an extensive historical perspective section by Hugo Fechtig et al. (2001) which described the early attempts and failures of dust detections until the grandiose achievements of the space missions to comet Halley and the beginning of the success story of infrared dust astronomy.

Progress in the last decade has been significant with sample return missions, in situ measurements and ground- and space-based measurements, as well as in laboratory experiments and modelling. Therefore, the time gap from the last comprehensive review and the pace of development merits a new coherent view in a conference and a book, now with an emphasis on interconnections, similarities, differences and on synthesizing results from different techniques. Astrobiological connections are a new aspect considered.

The first section of this chapter describes the turbulent early days of dust research that were focussed on the meteoroid hazard when the community painfully learned how to fly reliable in situ dust instruments. The second section characterizes the multitude of observational techniques contributing to our knowledge about cosmic dust. This section also describes the development of in situ instruments to complete dust telescopes and dust observatories in space by which dust astronomy is performed (Grün et al., 2005). Important laboratory experiments are discussed in section three. In section four we synthesize our understanding of the zodiacal cloud and give reference to modelling of other dust environments.

## 2. Dangerous Dust

Meteoroids are part of the space environment. The danger from millimetre and centimetre sized meteoroids to manned and unmanned activities was obvious from the beginning of the space age. Smaller particles could be dangerous to not well shielded satellites and to manned extravehicular activity as well. Therefore, at the beginning, all space-faring agencies started a vigorous program to characterize the meteoroid hazard as part of their space activity.

A set of simple microphones were the first dust detectors flown in 1950 in the US on board a V2 rocket. Parallel to the development of space dust detectors were appropriate simulation techniques. In order to simulate impacts and collision phenomena in space, accelerators were necessary to provide projectiles in the speed range at 10 km $s^{-1}$ and beyond. The work horse of accelerators for millimetre and bigger projectiles was the light gas gun. This gun consisted of a conventional powder gun that pushes a piston into a barrel containing hydrogen gas which itself accelerates the projectile. By this method a projectile speed up to 12 km $s^{-1}$ was reached, however the wear on the gun was excessive and, therefore, only speeds up to 5 km $s^{-1}$ were routinely achieved. A wide range of projectile materials can be accelerated with this method.

An electrostatic accelerator for micrometer sized and smaller particles is similar to nuclear physics devices that accelerate charged particles by high electric voltages (Friichtenicht 1962). Potentials of several million Volts are obtained by van-de-Graaff generators. High charges on electrically conducting particles are obtained by bringing these particles in contact with the fine tip of a tungsten needle which is at a high electric potential. In a 2 MV accelerator micrometer sized iron particles can reach speeds of 12 km $s^{-1}$, or 0.1 μm sized particles reach up to 35 km $s^{-1}$. Speeds as high as 100 km $s^{-1}$ have been observed for sub-micrometer sized grains. Electrically conducting particles consisting of iron, carbon, aluminum, and non-conducting glass, mineral, or plastic particles coated by conducting materials have been used in these accelerators. Currently, only the dust accelerator at the Colorado Center for Lunar Dust and Atmospheric Studies (CCLDAS, Thomas et al., 2013) is in routine operation (Fig. 2).

Many of the earliest dust detectors were single shot detectors responding to a single signal upon impact of a dust particle on to the sensor. These detectors had no means to directly distinguish dust impacts from any noise they responded to. Most popular were momentum sensors, consisting of a single piezoelectric crystal attached to an impact plate. Any signal exceeding a threshold was interpreted as an impacting meteoroid. Often such detector systems were calibrated by simply dropping glass beads on to them because they were not sensitive enough for projectiles from an electrostatic dust accelerator (Auer 2001).

In the 1960s, early microphone data from NASA's Explorer and some Russian Cosmos and Salyat satellites were interpreted as near-Earth dust enhancements by factors of 1000 and higher than the dust density in interplanetary space. Dust collections by primitive collectors on sounding rockets above the Earth's dense atmosphere seemed to confirm the satellite observations.

However, the early investigators did not appreciate that there were a variety of interfering effects that microphones respond to as well. These include external effects like cosmic rays hitting sensitive first stages of the amplifiers, or time variable ambient plasma effects that can interfere with sensitive dust detectors or temperature variations during a satellite's passage through the Earth's shadow causing thermal cracking. But also interferences generated inside the satellite like micro-vibrations caused by other spacecraft systems, or switching high currents or voltages could generate electromagnetic interferences in the dust detector. Likewise, pre- and post-flight contaminations of the collector surfaces invalidated the earlier in-flight rocket dust collections.

Measurements of the zodiacal light brightness set stringent constraints on the total dust density in interplanetary space. Theoretical analysis by Shapiro et al. (1966) and Colombo et al. (1966) discussed various aspects of 'The Earth dust belt: fact or fiction?' and concluded that a near-Earth enhancement of at most a factor of 10 could be reasonably explained but no enhancement factors of 100 or more. Indeed, the current best estimate of the near-Earth gravitational enhancement is only a factor of 2. At about the same time, Nilsson (1966) expressed doubts about the Earth's dust cloud based on experimental grounds. He used witness microphone detectors that were shielded form dust impacts but otherwise identical to active ones. They recorded about the same event rate as the active ones and hence, contradicted the dust belt hypothesis. Therefore, the initial high fluxes reported were erroneous and were caused by the combined effects of immature dust detectors and the harsh near-Earth environment.

The most reliable of the single parameter detectors were the penetration detectors on board several early satellites and space probes (e.g. Explorers 16 and 23, Pioneers 10 and 11, D'Aiutolo et al., 1967). These detectors consisted of a large number of pressurized cells (termed 'beer cans') that recorded the decrease in gas pressure when a wall was punctured by a meteoroid. The walls were metal sheets (stainless steel or copper beryllium) of 25 or 50 μm thickness and had penetration limits of $10^{-9}$ and $10^{-8}$ g at 20 km $s^{-1}$ impact speed (Humes et al., 1974). The Pegasus detectors were large area (about 200 $m^2$) detectors that recorded penetrations of up to 0.4 mm-thick aluminium sheets by the discharge of a capacitor. Detectors like these determined the flux of meteoroids in near-Earth space in the 10 micrometer to millimetre size range. This range was important for the assessment of the meteoroid hazard for typical satellites and the manned missions ahead. Cour-Palais (1969) published the official NASA meteoroid model that described the Earth's natural meteoroid environment. It showed that no special and demanding precautions had to be taken to protect space systems against the meteoroid hazard and that the Apollo missions to the Moon were safe to be conducted. It also showed that a large space station protected by an effective 1 cm thick aluminium shield could survive in the natural near-Earth meteoroid environment.

As a consequence, the interest in meteoroids shifted from technological to astrophysical questions and the amount of money to support dust studies dramatically declined. From early major groups at the NASA centres, only a few individual scientists continued to work in the meteoroid field, in Russia almost all meteoroid work subsided. Only a few universities and research laboratories, primarily the University of Kent at Canterbury and the Max-Planck-Institute for Nuclear Physics at Heidelberg, kept on advancing the meteoroid field.

Previous experience has shown that in situ measurements of space dust are low-level measurements, they require highest sensitivity for extremely scarce impact events in a hostile time-varying space environment. Only one dust impact per month is not uncommon, therefore, the foremost challenge was to develop means to distinguish dust impacts from noise. Careful characterization of the impact signal at a dust accelerator is a fundamental requirement. Multiple separate dust sensors, with one or more sensors shielded from dust impacts, provide some means of statistical distinction of impact events from noise events. Amplitude or even waveform measurements of the recorded signals can help directly to identify noise and dust signals. The most reliable method, however, is to record several signals from a single dust impact in coincidence. Multi-coincidence detectors based on the impact ionization process were successfully flown throughout the solar system and, together with appropriate data analysis, provided significant results about various dust environments (see below). A well-proven approach is to spend as much effort to understand the noise behaviour of a dust instrument in a specific environment as to analyze dust impact data.

Recently, the near-Earth dust environment has been getting back into the focus of space agencies. The ever increasing human space activity has lead to a hazardous condition that could jeopardize future space activities in low-Earth-orbit (below approx. 1,000 km altitude). Abandoned upper rocket stages exploded in space which had not completely used up their liquid fuel. Collisions of satellites over the Earth poles where many orbits of polar satellites cross each other shed large amounts of long-living debris into Earth orbit. Already in 1978, Kessler and Cour-Palais (1978) noted that the increased population of satellites will eventually lead to increased amounts of debris by accidental collisions. Shielding satellites by massive shields will eventually worsen the situation because even more mass needs to be transported to Earth orbit which will finally become debris. Kessler predicted that after 50 to 100 years a run-away process will be started and shatter all matter in low-Earth-orbit, leading to the formation of an artificial collisional ring until atmospheric drag will clear this region of space again.

Current measurements of the near-Earth debris by radar, lidar, and other methods (Fig. 3) indicate that it is already in or near the self-destruction mode, also evidenced by the frequent collision avoidance manoeuvres of the International Space Station (ISS) with trackable debris. Micrometeoroid or space debris impacts damaged some focal plane detectors of the XMM-Newton and Swift-XRT X-ray observatories (Carpenter et al., 2008). Clustering in the times of impacts is indicative of dust generation events. Such clusters have been reported by NASA’s Long Duration Exposure Facility (LDEF) IDE instrument (Oliver et al., 1993, Cooke et al 1993) and by ESA's DEBIE instrument (Schwanethal, et al., 2005). Currently, ESA estimates that almost one million space debris objects bigger than 1 cm orbit the Earth. Because of the disastrous effects of a collision of such a projectile with an active spacecraft, most space agencies adopted measures to reduce the detrimental effects of space flight. These measures include to empty liquid fuel from used rocket stages, to bring in-operational satellites to orbits where the risk of collisions is low or to "de-orbit" them into the Earth atmosphere. However, an ultimate solution to the space debris problem has not been found.

From 1984 to 1990, LDEF, was exposed to the space environment at about 450 km altitude. Studies of impact craters on returned LDEF surfaces, and a similar study by the European Eureca satellite, showed that the near-Earth meteoroid flux is about a factor of 2 higher than the flux in deep space because of gravitational concentration by the Earth (McDonnell and Gardener, 1998). Depending on the particle size, natural meteoroids are outnumbered by man-made space debris: below about 10 μm and above about 100 μm in size debris particles are more abundant than natural meteoroids. Therefore, monitoring the near-Earth dust environment is presently a routine activity of all major space agencies.

The meteoroid hazard in interplanetary space is currently not well understood, and human space travel beyond the Earth-Moon system will require higher attention in the future. A space station in Earth vicinity like the planned Deep Space Gateway in lunar orbit would be an excellent platform for future active or passive dust detection techniques. Pegasus-type in situ dust detectors may be needed to monitor the dust flux in interplanetary space between Venus and Mars. In addition, during the future human colonization of Mars, dust from the Martian satellites may play a similar hazardous role as space debris in the Earth environment. Interstellar travel by automatic probes to α Centauri and other near-by stars at 30,000 km $s^{-1}$ speed (0.1 speed of light) pose an unpredictable hazard: at such a high speed micrometer-sized interstellar particles carry the energy of rifle bullets!

# 3. Multifaceted Scientific Dust Observations

According to the 2017 IAU definition, meteoroids are solid natural objects of size smaller than roughly 1 meter moving in, or coming from, interplanetary space. Interplanetary dust (or micrometeoroids) is the small end (<~ 30 micrometer) of the meteoroid size distribution. However, there is a continuous distribution of objects up to the size of asteroids, comets and planetary satellites that are in intimate relation with and constitute the interplanetary meteoroid complex. There are several scientific methods to observe and analyze meteoroids and interplanetary dust (Koschny et al., 2019).

## 3.1. Large-scale distribution and composition of interplanetary and interstellar dust by remote observations

Since the 17th century, when Giovanni Cassini explained the relation between zodiacal light and dust in interplanetary space, zodiacal light observations have been the main tool to study space dust. However, observations from the ground are handicapped by weather effects, atmospheric dust and light pollution, and airglow. Such observations have the additional difficulty that the brightness along the line of sight (LOS) is generated from particles located at different heliocentric distances and observed at different scattering angles. Because the light intensity scattered at a certain angle strongly depends on particle size, such observations cannot separate the radial dust density from the size distribution. Only in the special case when observations are performed along the same LOS from two different positions in space can the spatial density of particles along the section between the two observation positions directly be derived (Dumont, 1983). With the photometers on board the Helios space probes Leinert et al. (1981) succeeded in the measurement of zodiacal light between 0.3 and 1 AU with the same scattering geometry. The authors determined the radial brightness profile ($\sim r^{-1.3}$) and found an inclination of 2° of the symmetry plane of the zodiacal cloud with respect to the ecliptic plane. From these observations Leinert and Grün (1990) concluded that the zodiacal cloud has a flattened lenticular shape with an axial ratio of 1:7.

Outside the Earth's orbit, zodiacal light was observed by photometers on board the Pioneer 10 and 11 spacecraft (Weinberg et al., 1974). The zodiacal light brightness was found to exceed the background out to 3.3 AU distance from the Sun (Hanner and Weinberg, 1973). On board Pioneer 10 and 11 there were two more dust instruments besides the zodiacal light photometer: the beer can penetration detectors (see below) and the Asteroid Meteoroid Detector (AMD, Soberman et al., 1974). The latter instrument consisted of four telescopes designed to detect scattered sunlight from individual meteoroids passing through their field-of-view. From the timing and the amplitudes of the received light signals, the trajectory of the meteoroid could in principle be reconstructed. Because of too low preset threshold sensitivity, however, the instrument was very noisy and no single trajectory could be uniquely identified (Auer, 1974). Due to this inability to determine meteoroid trajectories of the Pioneer AMD instrument, the proposed twin AMD instrument was removed from the preliminary Voyager payload. Given the unique launch window of Voyager (1976-1978) for the planetary grand tour to the four giant gas planets, no replacement was found in time and, hence, no dedicated dust instrument was onboard the Voyager spacecraft.

An interplanetary dust particle scatters only less than 10% of the incident sunlight and contributes to the zodiacal light. The rest of the absorbed energy (>90%) is re-emitted as thermal infrared radiation, mostly in the 10 to 50 μm mid-IR wavelength range. Because of this effect, zodiacal infrared emission is a much more prominent astronomical phenomenon than emission at visible wavelengths; however, infrared observations have to be performed from above the Earth's atmosphere.

A milestone in the observations of the local zodiacal dust disc was the first all-sky survey at infrared wavelengths, conducted by the Infrared Astronomical Satellite (IRAS) in 1983. It mapped the sky at 12, 25, 60 and 100 μm wavelengths. Besides the dominant foreground zodiacal emission, several hundred thousand sources were discovered including stars, galaxies, and solar system objects. Many stars with dust discs were identified through their infrared excess radiation; among them dust discs around Vega, β Pictoris, Fomalhaut, and ε Eridani (Blum, 2019). A surprise was the observed structure in the zodiacal cloud: IRAS discovered a complex system of asteroid bands and many comet trails (Eaton et al., 1984, Sykes et al., 1986). Later on, the COBE satellite obtained unprecedented information on the overall structure of the zodiacal cloud to an accuracy of better than 3% (Kelsall et al., 1998). Using the Spitzer infrared Space Telescope, Reach et al. (2007) found 80% of the Jupiter family comets to possess dust trails. An extreme example is comet 73P/Schwassmann-Wachmann (Fig. 4) that in 1995 started to break up into many large pieces which spread along the comet's original orbit. Recently, the Planck mission extended the observations of the zodiacal cloud to millimetre wavelengths. Given that IR observations from near the Earth are handicapped by the dense zodiacal foreground, a 'bird's eye view' from an IR observatory above the ecliptic plane (e.g. at 40° latitude) would be able to examine the extent and fine structure of the 'warm' zodiacal cloud, and, finally, may observe the 'cold' outer Trans-Neptunian dust disc. It may also provide an unobscured view into interstellar space at mid-IR wavelengths.

IR observations also provide compositional information about the dust particles that emit or absorb IR radiation. IR features are often the only way to obtain direct information on the composition of dust; this is especially true for interstellar and circumstellar space. Beside these compositionally identified features there are many prominent unidentified features in the dusty interstellar environment.

Amorphous silicates were the first solids discovered in circumstellar space. They were identified by their prominent broad features at 9.7 and 18 μm. Crystalline silicates such as pyroxenes and olivines have sharper features and were also commonly detected in the circumstellar environment (cf. Fig. 5). These features are very common in oxygen-rich asymptotic giant branch (AGB) stars. The most common carbon bearing solid in carbon-rich stars is SiC which has a feature at 11.3 μm. Finally, water ice has a feature at 2.05 μm.

Carbon bearing molecules display a number of features in the 3 to 11 μm range. They were identified in the spectra of planetary nebulae and came as a complete surprise. These features are generally identified as originating from stretching and bending modes of polycyclic aromatic hydrocarbons (PAHs), although it has recently been shown that they may also originate from amorphous hydrocarbons. Natural terrestrial counterparts of these compounds may be coal, kerogen and petroleum.

More complex hydrocarbons were identified in interstellar space. Interstellar dust grains with typical sizes in the sub-micrometer range are believed to consist of a silicate and/or carbonaceous nucleus with an ice mantle if the temperature is low enough.

In our solar system, since the exploration of comet 1P/Halley in the 1980s, it has been known that a significant fraction of the cometary organic matter exists in the solid phase and is ejected from the nucleus as dust particles. For example, phosphorus, a key element in living organisms, was recently identified in the gas phase of comet 67P/Churymov-Gerasimenko by the ROSINA instrument on board Rosetta (Altwegg et al., 2016).

The presence of organic molecules which may be relevant for the formation of more complex biogenic compounds in living organisms raises the question whether space dust may harbour the precursors of life (cf. Pintado et al., 2019).

Most planetary missions carried cameras on board to observe the target planet(s), satellites and rings. The TV cameras on board both Voyager spacecraft discovered new dusty phenomena around all giant planets (Smith et al., 1979a, 1979b, 1981, 1982, 1986, 1989). The jovian system harbours a complex ring system inside about 3 Jupiter radii ($R_J$) from Jupiter (Showalter et al. 2008, Hamilton & Krüger, 2008) and powerful volcanoes on Io that spew plumes of ash as high as 300 km above the surface (McEwen et al., 2004). At Saturn the Voyager cameras observed small inner satellites, interacting gravitationally with one another and with particles in the dense rings (A, B, C, and F rings). In the B ring the mysterious spokes phenomenon was discovered (Grün et al., 1983, Morfill et al., 1983, Goerz and Morfill 1983) that is still awaiting full explanation. Beyond the F ring two tenuous rings (G and E rings) were observed. At Uranus nine rings are described, along with the locations and sizes of 10 shepherd satellites. Like the rings of Uranus, those of Neptune are composed of very dark material. On the biggest geologically evolved satellite, Triton, at least two active plumes driven by sublimating nitrogen ice were found. Later the cameras on Galileo and Cassini followed up on further investigating the dusty phenomena in the Jovian and Saturnian systems. It was found that geysers on the small icy Saturnian moon Enceladus feed the broad dusty E ring.

### 3.2. Meteor Observations

On a clear night, about 1 to 10 faint meteors can be observed by the naked eye per hour. Meteors are caused by centimetre-sized and bigger meteoroids that enter the Earth's atmosphere at speeds greater than 10 km $s^{-1}$. Once in a while, when a meter-sized meteoroid hits the atmosphere, a bright fireball appears. More sensitive methods to observe and measure the trajectories of meteors are photographic and TV cameras and telescopes. Spectroscopy of meteors shows mostly lines of meteoritic metals (Na, Mg, Ca, and Fe, Vojacek et al., 2015), besides strong emission lines of atmospheric origin (O, N, and $N_2$).

While the speed of meteors can be accurately determined, the size or mass of the meteoroid can only be estimated. Calibration measurements of the luminous efficiency and ionization coefficient have been performed only with micrometer-sized particles at relevant entry speeds exceeding 10 km $s^{-1}$ (Friichtenicht and Becker, 1971, Thomas et al., 2016). For centimetre sized meteoroids only theories are available to obtain the meteoroid mass from the observed measurement (Koschny et al. 2019). Effects of meteors on planetary atmospheres like the formation of layers of smoke and ice particles and atmospheric gas-phase chemistry are discussed in Plane et al. (2019).

Even though our collections contain more than 10,000 meteorites, only sixteen falls of meteorites have been observed with adequate equipment to derive their interplanetary orbits. These meteorites are of the ordinary chondrite and achondrite types and their heliocentric orbits prior to entering the Earth's atmosphere had aphelia in the asteroid belt, directly indicating a genetic relationship between these meteorites and asteroids. A comprehensive review of meteor research was given by Ceplecha et al. (1998). However, all observational methods require a dark and clear sky and, therefore, they cannot provide a complete record of the Earth meteoroid environment.

About 80 years ago, radar techniques were developed to observe faint meteor trails even during daylight. Radar meteors can be detected using either high power and large aperture radars (like Arecibo or other big radar dishes) or specular backscattering meteor radars such as the Advanced Meteor Orbit Radar (AMOR) in New Zealand and the Canadian Meteor Orbit Radar (CMOR). The first type of radar observes the head-echo of the meteor which is the signal reflected from the plasma region travelling with the meteoroid. The detection of the head-echo allows for the determination of instantaneous meteor altitude, in-beam speed and deceleration. However, it suffers from the low head echo rates and from the low precision of the true velocity measurement. A specular meteor radar detects echoes from the ionized trail produced by meteors ablating in the Earth atmosphere. From the measurements of the same meteor trail by several transmitting and receiving stations on the ground, the velocity of the heliocentric orbit can be inferred.

Three major complete meteor surveys have been performed by averaging over daily and seasonal variations. The first study of apparent radiant concentrations used the Harvard Meteor Radar Project (HRMP, cf. Taylor and Elford, 1998). More recently, Galligan and Baggaley (2004, 2005) reported a meteor orbit study from AMOR. Highly sensitive AMOR observations indicate even an interstellar origin of a small percentage (<1%) of the observed meteors (Baggaley et al., 1993, cf. Hajdukova, 2016). In the Northern hemisphere the lower sensitivity CMOR (Brown et al., 2008) is operated. Besides the study of meteor streams, radar meteor data is used for the comprehensive analysis of the sporadic meteor background. While meteor streams show up in narrow (i.e. few degrees wide) concentrations of apparent radiants (not corrected for Earth motion), also sporadic meteors display wide and diffuse concentrations of radiants (Fig. 6). The symmetric north and south apex radiant concentrations correspond to meteoroids at the descending and ascending nodes of their inclined orbits. Similarly, the helion and antihelion radiants represent meteoroids on the outgoing and ingoing legs of their eccentric orbits. The radiant distributions are corrected for atmosphere observing biases which include correction for the radar's collecting area, the altitude dependent initial trail radius effect, Faraday rotation effects, the finite velocity effect and the pulse repetition factor (Campbell-Brown, 2008). The different radiant concentrations have different mean observed speeds and have been tentatively associated with different parent objects: helion and antihelion meteors with Jupiter family comets (~35 km $s^{-1}$), toroidal meteors with Halley group comets (~45 km $s^{-1}$), and apex meteors with long-period comets (~55 km $s^{-1}$, Jones et al., 2005).

### 3.3. Composition of meteorites

Football-sized and bigger meteoroids may survive the entry into the atmosphere, and some residual material may fall to the ground, where such meteorites can be picked up and examined. On the Antarctic continent almost any rock one finds on the blue ice surface is a meteorite fallen some time ago and transported to its finding site by ice motion.

Because of their exotic extraterrestrial nature, meteorites belong to the best analyzed rock samples on Earth. The most abundant meteorites are chondrites because they contain up to millimetre-sized round particles (chondrules) embedded in a fine-grained matrix (Fig. 7). These chodrules are composed mostly of silicate minerals, apparently being melted and solidified in space. Chondrites, especially carbonaceous chondrites (the matrix of which contains abundant carbonaceous material) consist of a mixture of elements believed to comprise the mean composition of the entire planetary system (called "cosmic abundances"), except for some volatile elements, like hydrogen and helium. This primitive material has not been modified by thermal processes contrary to the interior of planets. Less than 10% of the observed meteorite falls are stony meteorites (achondrites). They consist of material similar to terrestrial basalts or plutonic rocks that have been differentiated and reprocessed to a lesser or greater degree due to melting and re-crystallization on or within meteorite parent bodies. About 5% of the meteorites are iron meteorites consisting of almost pure iron-nickel. They are believed to originate from planetary cores. One percent of the meteorites are stony-iron meteorites (Pallasites) consisting of centimetre-sized olivine crystals in an iron-nickel matrix. While the chondrites represent the elemental composition of the terrestrial planets, the other meteorite classes (achondrites, pallasites, and iron meteorites) have been used as analogue materials for different compositional shells inside the Earth.

Mineralogical, chemical and isotopic analyses of meteorites have provided deep insights into the physical and chemical conditions during their formation, i.e. into the history of the planetary system as a whole. The asteroid belt has always been suspected to be a source of meteoroids. Most asteroids have orbits between Mars and Jupiter. Hirayama suggested that asteroids with very similar orbits form families that have been generated by the break-up of a larger asteroid in a catastrophic collision about 100 million years ago (Hirayama, 1923). Nesvorny et al (2003) identified the IR asteroid bands with asteroid families that were generated less than 10 million years ago. Also large impacts on Mars and on the Moon shed material into interplanetary space, several lunar and Martian meteorites were found on Earth.

Organic compounds identified in carbonaceous chondrites of aqueously altered petrographic types 1 and 2 include amines and amides; alcohols, aldehydes, and ketones; aliphatic and aromatic hydrocarbons; sulfonic and phosphonic acids; amino, hydroxy-carboxylic, and carboxylic acids; purines and pyrimidines; and kerogen-type material (Sephton 2002). An extensive range of amino acids was found in the Murchison CM2 chondrite. Organic solid materials in carbonaceous chondrites are dominantly an insoluble macromolecular component resembling terrestrial kerogen, which is the most common type of organic matter on Earth. The latter results from the decomposition of microbial and higher living tissue at the bottom of oceans or lakes, and it is used as a nutrient by bacteria. All known life is based on organic compounds and water; both of these are present in carbonaceous chondrites. Analyses of the organic matter in meteorites can provide insights into the types of chemical reactions and organic compounds which could have been significant on the prebiotic Earth (Pintado et al., 2019).

Up to now no meteorite has been linked to a cometary parent body. However, isotopic analyses of meteorites found that about $10^{-5}$ of the meteorite material has non-solar system isotopic abundance and is of pre-solar (interstellar) origin.

## 3.4. Structure and composition of collected cosmic dust particles

It is obvious that analyses of micrometeorites collected in the stratosphere could give similar information as meteorites. A high flying aircraft like the U2 spy plane can cruise at 20 km altitude for many hours. On its wings it carries a several 100 $cm^2$ flat plate dust collector which sweeps huge amounts of air providing a strong concentration effect on the collector.

Extraterrestrial grains (conveniently termed IDPs, which stands for Interplanetary Dust Particles) of about 5 to 50 μm in diameter are collected that way (Fig. 8). The lower size limit is due to contamination by smaller terrestrial particles. Micrometer and sub-micrometer-sized particles from volcanic eruptions can reach these altitudes in significant amounts. The upper limit is caused by the scarcity of bigger particles, e.g. only about ten IDPs of more than 10 μm in size are collected during one hour of aircraft flight. The extraterrestrial nature of the collected particles was demonstrated by their chemical composition which often resembles chondritic meteorites and the finding in IDPs of traces of solar wind helium and tracks from the exposure to high energetic ions in space (Brownlee et al., 1976). NASA has routinely performed cosmic dust collections by airplanes since 1981.

Like meteorites, IDPs do not represent an unbiased sample of meteoroids in space. Only small interplanetary dust particles of a few to 10 μm diameter are decelerated in the tenuous atmosphere above 100 km altitude, especially if their entry speed is low. At this altitude the deceleration is gentle, and the grains do not reach the temperature of substantial evaporation (T ≈ 800°C), because they effectively radiate away excessive heat. These decelerated dust particles subsequently sediment through the atmosphere and become accessible to collection in the atmosphere, on the Earth's surface and on the deep-sea floor. At 20 km altitude the concentration of interplanetary particles of 10 μm diameter is several thousand times higher than in space. Especially, the polar ice caps and mid-ocean sediments provide concentrations of ablated meteoritic material. Most of the micrometeorites collected in antarctic ice have primitive carbonaceous composition. Nesvorny et al. (2010) suggest that Jupiter family comets are the main source of zodiacal dust inside 5 AU and, thus, the most likely source of these micrometeorites.

In deep-sea sediments, cosmic dust deposits were already recognized in the 19$^{th}$ century (Murray and Renard, 1884). Recently, supernova-generated dust was identified on the ocean-floor, indicating that our solar system was hit by a nearby supernova shock wave that exploded ~50 pc away from our solar system 2-3 Myr ago (Knie et al., 2004, Fry et al., 2016). Extensive studies of the particle composition and variations in dust influx have also been performed (Prasad et al., 2017). Cosmic dust in the Earth atmosphere is discussed in detail in Plane et al., (2019). An influence of cosmic dust on the Earth climate is also being debated.

The first samples of extraterrestrial origin were the lunar samples returned by the American Apollo astronauts (380 kg) and by the automatic Russian Luna probes (ca. 0.3 kg). Analyses of the samples revealed their similarity with terrestrial material. Age determinations of lunar samples suggest that the Moon formed at approximately the same time as the Earth, perhaps by a giant impact on to the young Earth (Canup and Asphaug, 2001).

Recently, dust samples were collected and brought back to Earth from comet Wild 2 by the Stardust mission (Brownlee, 2014) and from asteroid Itokawa by the Hayabusa mission (Nakamura et al., 2011). In 2004 during a close (240 km) flyby of comet Wild 2 at a speed of 6 km $s^{-1}$ Stardust collected more than 10,000 cometary dust particles on its 0.1 $m^2$ aerogel collector and returned them to Earth (Brownlee et al., 2006). Among the Stardust collection of cometary particles there were those containing large crystalline Calcium Aluminum Inclusions (CAI, Fig. 8) formed close to the Sun at temperatures above 1300º C. They gave proof that large scale mixing occurred in the protoplanetary disc (Brownlee, 2014, Levasseur-Regourd et al., 2019).

In addition to the cometary particle collector, the Stardust spacecraft was equipped with a second aerogel collector specially designed for the collection of interstellar particles. It was exposed to the interstellar dust stream for about 200 days during the cruise phase of the spacecraft (Sterken et al., 2014). A total of seven particles were identified which are of potential interstellar origin (Westphal et al., 2014). The two largest particles contained crystalline silicates and, thus, may have been circumstellar condensates (Sterken et al., 2019). Alternative origins of the particles have not yet been finally eliminated, and definitive tests through isotopic analyses are necessary to prove their interstellar origin.

In 2005 the Japanese Hayabusa mission (Nakamura et al., 2011) touched down on asteroid Itokawa and picked up about 1500 dust particles from the surface and returned them to Earth. Mineral chemistry of the samples suggests that this ca. 0.5 km sized asteroid consists of pieces of the interior portions of a once larger asteroid. Dust from asteroid Itokawa proved to be the link between S-type asteroids and ordinary chondrites.

Future realization of current plans for sample return missions to asteroids, comets, Martian satellites, Trojans, and Galilean satellites will eventually provide further glimpses into the nature of early Solar System materials.

### 3.5. Size distribution of meteoroids obtained by microcrater studies

In 1969, when NASA's astronauts returned samples from the Moon, it was immediately recognized that the lunar rocks were peppered by micrometeoroid impacts (Fig. 9). Impact craters were identified on individual rocks from millimetre down to sub-micrometer in size. Several effects, however, made the analysis difficult. Since the exposure time of a given surface on a lunar rock could not reliably be determined, no absolute flux values could be derived but only relative values for different crater sizes. Morrison and Zinner (1977) concluded that only the steepest slope measured on a single surface was least vulnerable to effects from variations in exposure conditions and possible shielding by thin coatings of dust on the rocks. Calibration of microcrater dimensions with respect to meteoroid sizes was provided by Hörz et al. (1975) who found a size-dependent crater-to-projectile diameter ratio range of 2 to 9 for meteoroids of $10^{-18}$ to 1 g masses at an assumed average impact speed of 20 km $s^{-1}$. Grün et al., 1985 used these results and combined them with in situ measurements in order to derive the present microcrater production flux on the lunar surface. Studies of the effects of secondary microcraters produced by ejecta from primary impact craters on lunar samples showed that the number of microcraters is significantly increased for crater diameters below 10 μm. It was concluded that the interplanetary flux of meteoroids below $10^{-9}$ g is up to 2 orders of magnitude smaller than the ejecta flux on the lunar surface.

Since the return of lunar surface samples by the Apollo and Luna missions, the study of impact craters on material exposed to space (e.g. on the LDEF mission) has been used to characterize the flux of interplanetary micrometeoroids and man-made space debris particles (e.g. Love and Brownlee, 1995, and McDonnell et al., 1998).

### 3.6. Dust detections by in situ detectors throughout the planetary system

In situ dust detectors are carried by space probes throughout the solar system from within Mercury's orbit (Helios) to beyond Pluto's orbit (New Horizons) where no other method is capable to detect and analyze dust (Table 1). Based on the early negative experience to reliably detect dust impacts in near-Earth space, extra effort was spent to make dust detectors more reliable. Highly reliable impact momentum sensors were realized with Giotto's Dust Impact Detection System (DIDSY, McDonnell et al., 1987) in the dusty environment of Halley's comet. DIDSY employed various impact detection principles, most notably five piezoelectric crystals mounted at various positions of Giotto's 3 $m^2$ front and rear shields. Careful laboratory and theoretical calibrations (McDonnell et al., 1989) led to sensitivity maps of the front shield for coincident recordings by the impact sensors. For hypervelocity impacts of Halley particles on to the Giotto shield at 69 km $s^{-1}$ impact speed, the resultant crater ejecta enhance the effective impulse by factors of up to 10 (McDonnell et a1 1984). Piezo-electric sensors were also employed on the Dust Impact Monitor (DIM) onboard the lander Rosetta/Philae which landed on comet 67P/Churymov-Gerasimenko (Seidensticker et al., 2007, Krüger et al., 2015b). Modern microphone detectors like the Mercury dust monitor (MDM) onboard the BepiColombo/Mercury Magnetosphere Orbiter mission (MMO) use four highly sensitive piezoelectric ceramic sensor plates of 16 $cm^2$ each. The full waveform of any signal will be recorded and transmitted to Earth (Nogami et al., 2010). Calibration of electrostatic dust accelerators provides means to identify dust impacts, to measure the momentum transferred, and to estimate coarsely the impact speed from the signal form.

The penetration detectors on board the interplanetary Pioneer 10 and 11 spacecraft reliably recorded very low impact rates in the outer planetary system (Humes et al., 1980). During the passages through the dusty environments of Jupiter and Saturn a handful of penetrations were recorded but because of the instrumental dead-time of 87 minutes it was not clear whether more undetected penetrations had occurred. Actually, indeed, Pioneer 11 had passed close to Jupiter’s Gossamer ring which was discovered 5 years later in 1979 by the Voyager 1 space probe. A principal way to check the total number of hits received by the Pioneer 10 and 11 penetration detectors was to see at which number of recorded hits (initial number of 234 pressurized cells) no further counts were eventually recorded. However, this check became impossible because beyond approximately 20 AU the detectors responded to the very low ambient temperatures by freezing of the fill gas (argon and nitrogen, Humes, personal communication). The pressure in the cells dropped and the electrical discharge (which is used to detect the pressure drop) triggered. This discharge heated the gas and the pressure increased again, and after some time the whole process repeated. Because of this effect no useful data were obtained from the Pioneer 10 and 11 detectors beyond 20 AU. Nevertheless, these instruments characterized the meteoroid environment out to about 20 AU (Fig. 10) and confirmed that the outer solar system is safe to unmanned missions.

Despite the existence of reliable multi-coincidence dust detectors there was the desire to fly simple and inexpensive dust detectors on some space missions. On board the VEGA-l and 2 missions to comet Halley, Simpson et al (1986) used new light weight and low power consumption dust detectors. A PVDF detector consisted of a metal-coated and permanently polarized polyvinylidene fluoride film. A micrometeorite impact on to such a detector removes a portion of the metal surface layer and excavates some of the permanently polarized PVDF dielectric material underneath, thus generating a charge signal to determine the magnitude of the impact (Simpson and Tuzzolino, 1985, Shu et al., 2013).

Because of their low demand on spacecraft resources PVDF films were used as dust detectors on a number of missions. PVDF detectors provided useful dust information in the dusty environments of comets (Dust Flux Monitor Instrument on the Stardust mission to comet 81P/Wild 2, Tuzzolino et al., 2004) and planetary rings (High Rate Detector (HRD) of CDA on Cassini, Kempf, 2008). In the Earth environment, the SPADUS instrument (Tuzzolino et al., 2005) on the Advanced Research and Global Observation Satellite (ARGOS) used two arrays of PVDF dust sensors in a time-of-flight (TOF) arrangement that provided measurements of flux, velocity, and trajectory of big (~$10^{-9}$ kg) natural meteoroids as well as space debris particles. On the other hand, the measurements of small (~$10^{-12}$ kg) particles on the Cosmic Dust Experiment (CDE, Poppe et al., 2011) on the Aeronomy of Ice in the Mesosphere (AIM) mission suffered from high noise rates because PVDF displays both pyroelectric and piezoelectric properties and is affected by temperature variations and mechanical vibrations. Contrary to the SPADUS instrument, CDE could not rely on coincident signals from the same dust particle. The strong temperature contrast in sunlight and in the Earth's shadow caused both interferences from thermal cracking and from mismatch of electronic components.

On the New Horizons mission to Pluto and the Kuiper belt (Stern, 2008] the light-weight Venetia Burney Student Dust Counter (SDC, Horányi et al., 2008) was added at a late stage of spacecraft development. The instrument consists of fourteen independent PVDF film impact sensors of approximately 0.1 $m^2$ sensitive area each, twelve of which are exposed to space with the remaining two reference sensors shielded and placed on the underside of the instrument. The measured signal-to-noise ratio is approximately 0.5, hence, the measured meteoroid flux has at least a factor of two uncertainty (Fig. 10, Piquette et al., 2019). The large error bars are partially a result of the small number of events recorded, but are mostly the effects of a too small number of reference detectors. The number of counts and their errors on these detectors have six times the weight in comparison to the counts on the open detectors. This SDC data analysis demonstrates the importance for simple impact instruments like the PVDF detectors of having an independent characterization of noise. Without reference detectors the observed signals could be anything (noise event or dust impact), therefore such simple impact detectors require another independent means like reference detectors if applied in uncharted territory.

PVDF-type detectors were also used for the Arrayed Large-Area Dust Detectors in INterplanetary space (ALADDIN, sensor area 0.54 $m^2$) on board the Japanese Space Agency's (JAXA) IKAROS mission. This instrument measured interplanetary dust above 10 μm in size between Venus and Earth orbits (Yano et al., 2013, Hirai et al. 2014). Of 4427 events recorded within 300 days 60% had to be ruled out as noise (Hirai et al. 2017). The remaining 1773 events were classified as most promising dust impact events because they resulted in a dust flux compatible with the flux at 1 AU (Grün et al., 1985).

In the future, the Circum-Martian Dust Monitor (CMDM) will be a 1 $m^2$ thin-film sensor on board the Martian Moons Exploration (MMX) sample return mission to the Phobos and Deimos. CMDM will employ the outermost layer of the multi-layer thermal spacecraft insulation (MLI) together with piezo-electric sensors (Kobayashi et al., 2018a) to search for putative Martian dust rings (Soter 1971) which are still waiting for discovery (Showalter et al., 2006). An even larger (~3 $m^2$) PVDF detector will measure dust beyond the asteroid belt onboard the planned OKEANOS mission to the Jupiter trojans (Okada et al., 2018).

The only dedicated instruments on board the Voyager spacecraft to observe dust were the cameras which discovered and characterized the extensive ring systems around all four giant planets. However, there was a surprise dust instrument on board no one had foreseen. On August 26 1981, during the Voyager 2 Saturn flyby the plasma wave instrument recorded a very intense burst of noise spikes more than 100 times above the background in the few-minute interval around the ring plane crossing (Gurnett et al., 1983). The maximum intensity centered almost exactly on the ring plane crossing. The signals consisted of many brief impulses occurring at a rate of several hundred per second close to the newly discovered G-ring. The investigators interpreted the enhanced rate of spikes as impacts from a cloud of dust particles onto the spacecraft skin. During all subsequent flybys of the outer planets Uranus and Neptune this instrument recorded also concentrations of dust impact spikes in the corresponding ring systems. Similarly, in 1985 during a passage through the tail of Comet Giacobini-Zinner a plasma wave instrument onboard the NASA ICE spacecraft observed numerous dust impact spikes within a distance of 30,000 km from the comet nucleus (Gurnett et al., 1986).

More recently, the plasma WAVES instruments on board the interplanetary STEREO A and B spacecraft recorded two types of spikes on their electric field antennas (Meyer-Vernet, et al., 2009). The investigators suggested the voltage spikes on a single antenna to be due to a strongly varying flux of nanometer-sized dust particles arriving from the solar direction. These particles would be more than 10 times smaller than the earlier detected beta-meteoroids (see below). A second type of signals in the wideband waveform analyzer was based on coincident signals from all three antennas; these where identified to be due to impacts submicrometer and micrometer-sized dust particles, roughly compatible with classical interplanetary and interstellar grains. Laboratory dust accelerator calibration of plasma wave antennas has quantified the charge signals associated with hypervelocity dust impacts on materials specific to STEREO (Collette et al., 2015). Since 2009 the rate of single spikes has ceased on STEREO A but not on STEREO B. No such anomaly was observed on both spacecraft for the triple signals referring to the flux of micrometer sized grains (Malaspina et al., 2015), and no explanation for the disappearance of nanometer dust spikes on STEREO-A (but not STEREO-B) was found either. Recently, the interpretation of the single hits as dust signals was put into question (Kellogg et al., 2018). An independent confirmation of the interplanetary flux of nanometer-sized dust particles from the solar direction is still missing.

Dust instruments in combination with solar wind spectrometers and magnetometers may give new information on how plasma-dust interactions affect solar wind dynamics and composition. A nano-dust analyzer (O'Brien et al., 2014) positioned at 1 AU and observing the flow of dust particles will be a suitable instrument to determine the formation, dynamics, and composition of nano-dust particles in the inner solar system (Juhasz and Horanyi, 2013). Unfortunately, the Parker Solar Probe mission does not carry such a dedicated dust instrument when it will pass through the densest region of the zodiacal cloud in the solar F-corona within 10 solar radii (0.05 AU) from the Sun with a speed of almost 200 km $s^{-1}$ (Mann et al., 2004). The plasma wave antennae as well as many other instruments and spacecraft systems on board may be affected by impinging high-speed dust particles.

The Cassini spacecraft orbiting Saturn from 2004 to 2017 carried both a dedicated Cosmic Dust Analyzer (CDA, Srama et al., 2004) and a Radio and Plasma Wave Science instrument (RPWS Gurnett et al., 2004). In this case both methods could be directly compared and the relative sensitivities were determined (Ye, et al., 2014).

The most successful and versatile impact phenomenon used in dust detection and analysis is impact ionization. This phenomenon had been theoretically predicted by Yu. P. Raizer (1960). Shortly after the first dust accelerator was developed at the TRW Company Friichtenicht and Slattery (1963) reported the experimental verification of impact ionization. An impact of a fast (> 1 km $s^{-1}$) particle on to a solid target generates an expanding impact plasma cloud. In an impact ionization detector ions and electrons are separated by an external electric field and two coincident and complementary charge signals are recorded that provide a very sensitive and reliable method of dust detection. The early impact ionization detectors flown on the OGO satellites and on the lunar Explorer 35 (Alexander et al., 1971) were time-of-flight (TOF) systems consisting of a thin film front sensor and a rear sensor 0.1 m apart. Both the dust penetration of the front film and the impact on to the solid rear sensor generated coincident impact charge signals. However, mislead by the early false reports of a very high micro-particle flux in the Earth' environment, Alexander selected only 0.0005 $m^2$ as sensitive area of a single detector. As a consequence no impacts were recorded.

Berg and Richardson (1969) extended this idea by combining 16 TOF tubes into single 0.01 $m^2$ dust detectors that successfully operated for more that 7 years on the interplanetary space probes Pioneer 8 and 9. These detectors made the first important observations of interplanetary dust and discovered a flow of particles on hyperbolic orbits arriving roughly from the solar direction (Berg and Grün, 1973). This outward dust flow was interpreted by Zook and Berg (1975) as small grains being generated by collisions of meteoroids near the Sun and being expelled by the prevailing action of solar radiation pressure. Because the ratio of solar radiation pressure force $F_{rad}$ and solar gravity $F_{grav}$ is often called $\beta = F_{rad}/F_{grav}$, the authors coined the name beta-meteoroids for these particles. For particles to be affected by radiation pressure beta has to be large: A particle with $\beta \geq 0.5$ released from a parent body on a circular heliocentric orbit will leave the solar system on a hyperbolic trajectory. Since their discovery beta-meteoroids have been observed by dust instruments on Helios 1 (Grün et al., 1980) and Ulysses (Baguhl et al., 1995, Wehry et al., 2004).

In 1968 the European Space Research Organization ESRO - the precursor organization of ESA – offered the opportunity to fly an instrument on the HEOS 2 satellite. The Max-Planck-Institute for Nuclear Physics proposed a new dust detector and was selected. From laboratory experiments (Grün and Rauser, 1969) it was known that a film has a strong effect on projectiles that penetrate it at high speeds even if the film is much thinner than the diameter of the projectile. Therefore, a film-less impact ionization dust detector was flown on the HEOS-2 satellite (Dietzel et al., 1973, Hofmann et al., 1975). It consisted of a hemispherical impact target of about 0.01 $m^2$ sensitive area and a central ion collector. A voltage of 350 V between the target and the central electrode separated the impact charges, and two coincident signals were recorded upon impact of a fast dust particle on to the target. Within 2.5 years more than 400 particles were recorded on the highly eccentric orbit between 10,000 and 240,000 km above the Earth's surface (Fechtig et al., 1979). At large distances (>60,000 km) the observed flux of interplanetary dust was mostly randomly distributed in time. Groups of several particles within a few hours were interpreted as members of ejecta clouds generated by large meteoroid impacts on the Moon. Swarms of impacts were recorded in the auroral zone of the Earth's magnetic field and in the perigee region of the spacecraft orbit. Electrostatic disruption of large porous meteoroids has been suggested as a mechanism of these bursts. These latter phenomena await confirmation by modern dust instruments in the Earth-Moon system.

A major improvement of the simple HEOS dust detector configuration was a ten times increase of sensitive area (0.1 $m^2$). Based on the high reliability of impact detection, a channeltron ion detector was introduced inside the central ion collector, thereby providing a third coincident signal upon dust impact. Additionally, variable commandable detection thresholds and a variable coincidence scheme controlled by a programmable processor were implemented. Such a detector was successfully operated on board Galileo for 6 years en route to Jupiter and for about 8 years in the harsh Jovian environment (Grün et al., 1992a, Krüger et al., 2010). The twin detector on the Ulysses mission was operated in interplanetary space for 17 years (Grün et al., 1992b, Krüger et al., 2015a). In a noisy environment, multi-coincidence dust detectors like the ones on Galileo, Ulysses and recently the LDEX detector on LADEE (Horanyi et al., 2014) have the advantage that they provide means of identification of even unexpected noise sources (cf. Baguhl et al. 1993). Thereby, data analysis methods can be developed that reliably distinguish noise from dust impact events. By reprogramming Galileo's onboard data processing computer, this detector reached the same sensitivity and reliability as the Ulysses detector despite the fact that Galileo's data transmission rate to the ground was lower by more than a factor of hundred due to a spacecraft antenna failure (Grün et al. 1995).

Among the significant discoveries by the Galileo and Ulysses dust instruments were: the detection of streams of nanometer sized dust particles from the volcanoes of Jupiter's moon Io (Grün et al., 1993, Horanyi et al., 1993, Zook et al., 1996, Graps et al., 2000, Krüger et al., 2006, Hillier et al., 2019), impact-generated dust clouds surrounding the Galilean moons and the formation of a tenuous dust ring (Krüger et al., 2003, Krivov et al., 2002), and dust in and beyond Jupiter's gossamer rings (Krüger et al., 2009). By a Jupiter flyby, the Ulysses spacecraft was brought on to an orbit almost perpendicular to the ecliptic plane (Fig. 11). This provided the basis to probe the three-dimensional structure of the zodiacal dust cloud (Grün et al., 1997). The delay of the Ulysses launch by about 4 years and the subsequent Jupiter flyby caused the final orbit plane to rotate by about 90° compared to the originally planned launch date, which enabled the detection of an interstellar dust flow through the planetary system (Grün et al., 1994, Krüger et al., 2007, 2015, Strub et al., 2015, 2018). With the original launch date interstellar dust particles would not have entered the Ulysses dust sensor. Through the 8-year long measurements by Galileo the dust environment in the Jovian system is better understood than the natural meteoroid environment in the Earth-Moon system. The spare unit of the Ulysses dust instrument GORID (Geostationary Orbit Impact Detector) was flown on the Russian EXPRESS spacecraft and characterized the natural and space debris environment at ~36,000 km altitude (Drolshagen eta al., 1997). Only recently, measurements by the dust detector on the lunar orbiter LADEE (Lunar Atmosphere and Dust Environment Explorer, 2013) identified also an impact induced dust cloud around our Moon. The Lunar Dust Experiment (LDEX) instrument observed particles ejected by impacts of sporadic meteoroids and from meteoroid streams on to the lunar surface (Szalay and Horanyi, 2016a, b).

### 3.7. Composition measurements by dust analyzers

Dust instruments that characterize more than merely dust impacts and their magnitudes are called dust analyzers. The chemical and isotopic compositions and physical properties of dust particles can provide insights into the physical and chemical conditions during solar system formation and into the history of the planetary system as a whole.

The first dust analyzers used in space were compositional analyzers. Early laboratory studies of impact ionization already used a TOF impact mass spectrometer to analyze the ions released during a hypervelocity impact (Auer and Sitte, 1968, Hansen, 1968). The mass resolutions of the spectra obtained in the laboratory were low ($M/\Delta M \sim 30$), mostly because of electronic limitations. The first low mass-resolution dust analyzer was flown on Helios 1, 2 in 1974/76 (Dietzel et al., 1973). This dust analyzer consisted of a venetian blind type impact target behind which the ion collector was situated at the end of a one meter long tube. This TOF ion mass analyzer had only a marginal mass resolution of 5 – 10. In interplanetary space it identified three types of spectra dominated by low mass (<30 amu), medium mass (30-50 amu) and high mass (>50 amu) species (Altobelli et al., 2006). However, the main value of this mass analyzer was that each recorded ion spectrum reliably marked a dust impact. This feature was especially important since the noise rate was high. Helios 1 recorded 235 impacts during its 10 orbits about the Sun between 0.3 and 1 AU (Grün et al., 1980). Each of the spinning Helios spacecraft carried two dust analyzers of 0.006 $m^2$ sensitive areas each: the ecliptic sensor that scanned through the Sun direction was covered by a thin film, and the open south sensor on Helios 1 was shielded from direct sunlight by the spacecraft frame. Although, both sensors had widely overlapping fields-of-view the south sensor recorded significantly higher impact rates at otherwise similar impact parameters (impact direction and impact charge). A comparison with penetration studies showed that particles which did not penetrate the entrance film must have had bulk densities below 1000 kg $m^{-3}$. Approximately 30% of the particles on high eccentricity orbits ($e > 0.4$) had such low densities (Grün et al., 1980, Pailer et al., 1980).

A breakthrough was reached with the dust analyzers on the Halley missions, Giotto, VeGa 1, and 2 (Kissel et al., 1986). By inclusion of an electrostatic reflectron (Mamyrin et al., 1973) and faster electronics in the TOF spectrometer the mass resolution was improved to $M/\Delta M \sim 100$. Because of the dusty cometary environment these instruments needed only 5 $10^{-4}$ $m^2$ sensitive area. The dust impact mass spectrometers PIA and PUMA on the GIOTTO and VEGA spacecraft, respectively, allowed the compositional analysis of individual particles (Kissel et al. 1986a, b). The measurements showed that each particle is an intimate mixture of a mineral core, ices, and organic molecules. Since the impact velocity was large (>60 km $s^{-1}$) only atomic ions were identified in the Halley case. Nevertheless, Kissel and Krueger (1987) found evidence for the organic nature of cometary material. Important results were the discovery of a prominent carbonaceous component in Halley dust, and a wide scatter in some isotopic ratios. Cometary particulates are an intimate mixture of two end-member components, namely refractory carbonaceous (rich in the elements H, C, N, and O) and stony material (rich in rock-forming elements as Si, Mg, Fe), respectively. The stony component comprises silicates, metals, oxides, sulfides and others. Both end-member components do not occur as pure components but are mixed to the finest scale. Magnesium isotope ratios showed only a slight variation around the nominal solar value, whereas the isotopic ratio of $^{12}C/^{13}C$ showed large variations from grain to grain, but on average it was also solar like (Jessberger and Kissel, 1991). The average elemental composition was found to be solar like, but significantly enriched in volatile elements H, C, N, and O compared to CI chondrites (Jessberger et al., 1987).

A 20 times enlarged version ($10^{-2}$ $m^2$ sensitive area) of the Halley mass analyzer with mass resolution $M/\Delta M \sim 100$ was flown on the Stardust mission to comet Wild 2 (Kissel et al., 2004). For the first time a series of positive and negative ion mass spectra from the impact of (apparently) interstellar and cometary dust particles has been collected. In the spectra of 45 presumably interstellar particles, quinone derivates were identified as constituents in the organic component. 29 spectra obtained during the flyby of Comet 81P/Wild 2 on 2 January 2004 suggest the predominance of organic matter (Kissel et al., 2005). The authors found that organic material in cometary dust seems to have lost most of its hydrogen and oxygen as water and carbon monoxide. These are now present in the comet as gas phases, whereas the dust is rich in nitrogen-containing species.

The Cassini CDA instrument which was developed by Srama et al. (2004) is a combination of the Galileo type dust detector with a linear impact ionization mass analyzer of mass resolution $M/\Delta M \sim 30$ and 0.016 $m^2$ sensitive area (Fig. 12). The impact ionization detector and compositional analyzer is combined with a PVDF High Rate Detector (HRD) in order to time-resolve the flux of micrometer-sized particles during the fast passage of denser ring regions of Saturn. In contrast to the Galileo and Ulysses dust detectors which were mounted on a spinning spacecraft (Ulysses) or the spinning spacecraft section in the case Galileo, CDA was on board the 3-axis stabilized Cassini spacecraft. In order to provide some control of the pointing directions, CDA was mounted on an instrument controlled turn-table. Another significant feature during the initial dust measurements was reduced count rate data transmitted to ground through the continuous spacecraft engineering data stream while high resolution raw data and spectral data was included in the sporadic science data stream. This way CDA took dust measurements during Cassini's 7 year long trip through the planetary system to Saturn. Handicapped by operational constraints, CDA was able to obtain only two spectra of interplanetary particles inside Jupiter's orbit (Hillier et al., 2007). Their iron-rich composition with the lack of silicates and magnesium is in strong contrast to the silicate-rich minerals in interplanetary dust particles collected in the Earth's stratosphere. From hundreds of mass spectra obtained during Jupiter flyby CDA confirmed the compositional relation of dust stream particles with the volcanic plumes on Io. During 13 years in orbit about Saturn, CDA characterized thousands of mass spectra of particles in Saturn's E-ring and from the interior of Enceladus (Fig. 13). Postberg et al. (2011) conclude that a salt-water reservoir is the source of a compositionally stratified plume on Enceladus (see also Hsu et al., 2015, Hillier et al., 2019). CDA also identified particles above the main rings (Hsu et al., 2018), dust from the Kuiper belt and interstellar dust grains (Altobelli et al. 2016).

The Rosetta high resolution time-of-flight ion mass spectrometer for the analysis of cometary particles, COSIMA, included a dust collector, a manipulator unit for target handling, and a dust analysis station. More than 10,000 particles were collected which range from 50 to more than 500 μm in size. Most collected particles originate from the disruption of large aggregates (> 1 mm in size) during the collection process (Langevin et al., 2016, Merouane et al., 2016). The particles are composed of various minerals (silicates, Fe sulfides, etc., Hilchenbach et al., 2016). Fray et al. (2016) report the detection by COSIMA of solid organic matter in the dust particles emitted by comet 67P/Churyumov–Gerasimenko; the carbon in the organic material is bound in very large macromolecular compounds, analogous to the insoluble organic matter found in carbonaceous chondrite meteorites.

The cometary sampling and composition (COSAC) experiment on board the Rosetta lander Philae reported the detection of volatile organic molecules that were most likely released from dust particles collected in the instrument's entrance tube when Philae bounced at the cometary surface. The identified compounds are consistent with the first steps of photochemical and/or radiolytic evolution in ices—steps that could have taken place before the comet formed, in the protosolar molecular cloud or on the edge of the solar nebula (Goesmann et al. 2015). One of the key objectives of the Philae mission was the in situ analysis of a fresh sample of cometary material drilled from underneath the surface. This experiment failed, however, due to Philae's unfavourable landing conditions at the Abydos landing site. Keeping in mind that comets contain the most pristine material from the early solar system we know of, in situ sample analysis of fresh sub-surface cometary material in combination with cryogenic sample return should be given the highest priority for future cometary missions.

Recently, noncontact methods to analyze dust particles became available and provide means to derive the trajectories of dust particles in space. The combination of a trajectory sensor with a large area compositional analyser is called a Dust Telescope (Fig. 14, Srama et al., 2005b, Sternovsky et al., 2011). It measures the trajectory together with the composition of dust particles entering the instrument. In the near future the Japanese-German collaboration in the Destiny+ mission will employ such a dust telescope (Kobayashi et al., 2018b). This mission will focus on analyses of dust populations at 1 AU and the contributions from asteroid (3200) Phaethon which is the parent object of the Geminid meteor stream. The dust telescope will be able to distinguish lunar ejecta particles from cometary and asteroidal interplanetary and interstellar dust particles and to characterize their compositional similarities and differences. The measurements will provide a direct comparison of the composition of interstellar raw material with the more processed material from comets and asteroids. Such analyses will complement currently planned asteroid sample return missions and pave the way for future interstellar and interplanetary dust collection missions from near-Earth space (Strub et al., 2019).

A dust telecope of similar design will be on board NASA's Europa Clipper mission to Jupiter's moon Europa. The SUrface Dust Mass Analyzer (SUDA) will measure the composition of ballistic dust particles populating the thin exospheres that were detected around each of the Galilean moons. Since these particles are direct samples from the moons' icy surfaces, unique composition data will be obtained that will help to define and constrain the geological activities on and below the moons' surface. SUDA will characterise the surface composition, habitability, the icy crust, and exchange processes with the deeper interior of the Jovian icy moon Europa (Kempf, 2018).

### 3.8. Dust observatories in space

Comets have long been recognized as carriers of pristine, unheated and unaltered material from the early formation stages of our planetary system and at the same time to be significant sources of dust in the solar system. Therefore, the interest in studying these solid, icy, and gaseous objects has been longstanding. Cometary dust observatory missions, generally, have dust science as one of their prime mission objectives. Therefore, they carry advanced dust analyzers together with a full set of other environmental monitors. Examples are the Halley missions, Giotto and VeGa, Stardust to comet Wild 2, and Rosetta to comet Churyumov-Gerasimenko (Levasseur-Regourd et al., 2019). Each of these missions carried several dedicated instruments to pursue various aspects of dust science (Table 2).

Because of the high eccentricity of their orbits, comets are difficult to reach and to rendezvous with from Earth. Cheaper ways are flyby missions that just cross the orbit of a comet close to the nucleus. In 1986 the Russian VeGa 1 and 2 and ESA's Giotto missions were the first missions to make close observations of Halley's comet. The VeGa missions passed by the nucleus at about 9000 km and Giotto at 600 km distance. During their high-speed flybys at about 70 km $s^{-1}$ relative to the comet they took images of the nucleus, determined its optical properties, and characterized the dust emission, the neutral gas and plasma environments.

The three space probes carried a complement of specifically designed dust instruments close to the comet. Most of these instruments were impact ionization detectors and mass spectrometers, but also piezo-electric microphones, and thin film detectors were flown. Important results were the discovery of (1) both, smaller and bigger grains in the coma than the 1 to 10 µm sized grains that had been anticipated from astronomical observations (McDonnell et al., 1987, 1989, Mazets et al., 1987), (2) the existence of a significant carbonaceous component in Halley dust (Kissel and Krueger, 1987), and (3) a much wider scatter in some isotopic ratios (Solc et al., 1987) than has been found in any other extraterrestrial material.

Analysis of observations by the Halley Multicolor Camera of the dust distribution near the nucleus of the Halley comet (Keller et al., 1990) revealed the occurrence of particle fragmentation close to the comet's nucleus. This suggested a sublimation process in which only a part of the gas comes directly from the surface, while a fraction is liberated during the fragmentation of dust particles. Such a distributed source of CO has been directly observed by the Giotto neutral gas mass spectrometer (NMS, Eberhardt et al., 1987) in the inner coma less than 20,000 km from the nucleus. It was proposed that the CO or a very short-lived parent of this molecule was released in the coma from cometary dust grains, such as the carbon, hydrogen, oxygen, and nitrogen rich "CHON" particles.

In 1992 during Giotto's extended mission the spacecraft flew by comet Grigg-Skjellerup at a distance of about 200 km from the nucleus. During the encounter, several meteoroid impacts were detected on Giotto's front shield (McDonnell, et al., 1992). The particle masses were found to exceed $10^{-8}$ kg, suggesting that the mass distribution of the cometary dust was dominated by relatively large particles. The results indicate a higher rate of mass loss from the nucleus than previously thought, and hence a higher dust-to-gas mass ratio.

The cosmochemical relation of cometary dust to meteorites and IDPs was in the focus of the Stardust mission. In 2004 it passed by comet Wild 2 within 230 km at a relative speed of 6 km $s^{-1}$ (Brownlee at al., 2006). The spacecraft carried the 0.1 $m^2$ Aerogel Dust Collectors, a dust spectrometer (Cometary and Interstellar Dust Analyzer, CIDA), and a PVDF Dust Flux Monitor Instrument (DFMI). During the flyby, DFMI observed a highly varying dust flux on small spatial scales, which was explained by the presence of jets and by fragmentation (Green et al., 2004). The mass of the dust coma was dominated by larger particles (up to millimetre sizes), as was found for comets Halley and Grigg-Skjellerup.

The Stardust spacecraft collected thousands of particles from comet 81P/Wild 2 and returned them to Earth for laboratory study. Upon collection, these particles produced hypervelocity impact features on the aerogel collector surfaces. The morphologies of the deceleration tracks were created by particles varying from dense mineral grains to loosely bound, polymineralic aggregates, ranging from tens of nanometres to hundreds of micrometers in size (Hörz et al., 2006). Some particles larger than a few micrometers were seemingly undamaged, whereas smaller or finer-grained components were severely altered. A major portion of the collected particles larger than a micrometer was composed of the silicate minerals olivine and pyroxene (Brownlee et al., 2006). For example, x-ray spectral analysis of a single particle showed three major components: sulfide pyrrhotite, enstatite, and fine-grained porous aggregate material with approximately chondritic elemental composition. Another particle as well as many of the other fragments are isotopically and mineralogically linked to CAIs which are exotic refractory components well-known from primitive meteorites that appear to have formed in the inner regions of the solar nebula. In addition to silicates and abundant sulfides, the collected comet samples contain organic materials. However, potential organic contaminants in the aerogel collector have greatly complicated the interpretation of the organic portions of the samples returned by the Stardust spacecraft (Sandford et al., 2010).

The comet samples returned by the Stardust mission showed that the cold regions of the early Solar System were not isolated but that the comets formed from mixed materials, including familiar high-temperature meteoritic materials, such as chondrule fragments that were transported to cold nebular regions (Brownlee, 2014). Interstellar dust collections by Stardust have been discussed above (Sterken et al., 2019).

After ten years of interplanetary cruise and 3 flybys at Earth and at Mars the Rosetta spacecraft reached comet Churyumov-Gerasimenko in August 2014. For more than 2 years Rosetta followed the comet from 3.6 AU inbound, through perihelion at 1.24 AU, to 3.6 AU outbound. During that time it orbited the approximately 2 km radius nucleus from about 10 km to several hundred kilometres distance until September 2016 when the mission was terminated by disposing the spacecraft on to the comet's surface. In November 2014 Rosetta delivered the Philae Lander to the nucleus surface where it took measurements of the nucleus properties at the Agilkia and Abydos landing sites. A synthesis of the morphology of dust observations at comet 67P was recently given by Güttler et al. (2019), and a comprehensive review of the science results from the observations of the various Rosetta instruments is found in Levasseur-Regourd et al. (2019). Here, we stress the complementary nature of the dust observations by various instruments.

The GIADA dust instrument on board Rosetta (Colangeli et al., 2007) was a combination of the Grain Detection System (GDS) and the Impact Sensor (IS).

GDS consisted of a light curtain (of 100 $cm^2$ area) generated by four laser diodes to illuminate passing dust particles so that they could be sensed by a series of photodiodes. The Impact Sensor was located 10 cm below GDS and consisted of a square aluminium plate (sensitive area 100 $cm^2$) equipped with five piezoelectric sensors. From the combination of signals from both systems the speed and the momentum of cometary grains were determined (Della Corte et al., 2015). Recordings by the IS alone provided just the momentum of the grains. GDS-only signals were obtained and were interpreted as very low density (~1 $kg/m^3$) dust particle impacts which did not trigger the more sensitive IS detector (Fulle et al., 2015). Independent confirmation of the existence of such particles is still pending.

One of the prime objectives of the Rosetta mission was to develop an understanding of cometary activity. Early in the mission, a plan was developed to implement an activity campaign involving most orbiter instruments in coordinated observations of an active area at the surface and the resulting dust jet above. The most critical aspect was a fly-through the jet by Rosetta to measure in situ gas (by the ROSINA instrument) and dust flux (GIADA) and to determine the chemical and physical dust properties (COSIMA, MIDAS) in and outside the jet. These measurements were to be compared with long-term (several nucleus rotations) remote sensing observations of optical, near-infrared, and thermal properties (OSIRIS, VIRTIS, MIRO) of the active region and its surroundings. Long-term limb observations (ALICE, MIRO, OSIRIS, VIRTIS) just above the nucleus' surface of the associated jet were planned to complement the campaign. Such observations were foreseen for mid 2015; however, an unexpected phenomenon terminated all plans for an activity campaign: the star sensors onboard Rosetta became confused by thousands of big dust particles in their field-of-view and could no longer support the attitude determination of the spacecraft. As a consequence the spacecraft had to retreat to several 100 km distance during perihelion passage of the comet where in situ observations of gas and dust became impossible and only the camera OSIRIS could observe a firework of outbursts during 3 months around perihelion (Vincent et al., 2016; Fig. 15).

Fortunately, in the last year of the mission, when the Rosetta spacecraft could approach the nucleus again within 30 km distance, three outbursts were serendipitously observed by several in situ and remote sensing instruments. On 19

February 2016, at a distance of 2.4 AU, nine Rosetta instruments observed an outburst of gas and dust from the nucleus of comet 67P/Churyumov-Gerasimenko. Among these instruments were cameras and spectrometers ranging from UV over visible to microwave wavelengths, in situ gas, dust and plasma instruments, and one dust collector (Grün et al., 2016). Furthermore, on 3 July 2016, six Rosetta instruments detected an outburst of gas, dust and ice particles at 3.3 AU from the Sun and took images of the outburst site (Agarwal et al., 2017). The largest number of dust particles in an outburst was recorded by GIADA on 5 September 2016 at 3.6 AU (Della Corte et al., 2017). Gas, dust, and ice particles were simultaneously recorded by ROSINA and the star sensors, and an image of the plume was obtained by OSIRIS. The combination of observations from various instruments bears great potential for unravelling the secrets of the comet.

Voyager fly-bys at the outer planets Jupiter, Saturn, Uranus, and Neptune demonstrated that dust is a major constituent of these planets' environments. There is an intimate interrelation between the satellites, rings, and magnetospheres. The Galileo mission visited the Jupiter environment with a simple dust detector, cameras, plasma and fields instruments. However, Saturn was the first planetary system that was explored by a dust observatory. The central dust instrument was the Cassini Dust Analyzer (CDA) that had capabilities to measure the composition and charge state of detected particles (Hsu et al., 2018). Together with the cameras and the full range of particles and fields instruments, Cassini characterized the interrelations between Saturn's satellites, magnetosphere and rings. Using Cassini nano-dust stream measurements, it was possible to derive the interplanetary magnetic field structure during the 2013 Saturn aurora campaign (Hsu et al., 2013, 2016). Due to the complex dynamical interactions with the interplanetary magnetic field, a fraction of fast nanodust particles emerging from the Saturnian system was sent back into the magnetosphere and could be detected by a spacecraft located within. It was demonstrated by CDA compositional measurements that Enceladus' particles probe the deep interior conditions of this satellite (Postberg et al., 2011, Hsu et al., 2015, 2018).

# 4. Laboratory Experiments with Dust

Important elements of dust science are laboratory experiments for the study of dust properties and processes. Such experiments address the formation, release from parent bodies, and destruction of dust and its interaction with its environment. We summarize here the most important simulation experiments that had an impact on the development of the field in the past. Recently, many more laboratory experiments have been developed to deepen our understanding of the interrelation between dust and its environment in space.

## 4.1. High speed collisions

The importance of dust accelerator experiments for the development and calibration of dust detectors has been already mentioned. Because of the limited number of projectile compositions used in the past, the need for an increased number of different projectile materials was recognized (cf. Hillier et al., 2014). Similarly, the study of craters on planetary and man-made collector surfaces benefits from impact experiments (cf. Hörz et al., 1975). Impacts on to ices (Timmermann and Grün, 1991, Koschny and Grün, 2001) support studies of planetary processes in the outer planetary system where ices are abundant. Dust accelerator experiments simulating the entry of meteors into the atmosphere provide important calibration of the ionization yield for radar meteor studies (Thomas et al., 2016).

An important parameter to estimate the life time of a meteoroid against mutual collisions is $\Gamma(u) = M_T/m_{pc}$, where $m_{pc}$ is the mass of the smallest projectile that at impact speed $u$ is able to catastrophically shatter the target object of mass $M_T$. The parameter $\Gamma$ is related to the threshold energy density $Q_s$ (J kg$^{-1}$) required to shatter the target: $Q_s = u^2 /(2\ \Gamma)$. $Q_s$ is defined as the ratio of projectile kinetic energy to target mass needed to produce the largest intact fragment of mass $M_L$ that contains half of the target mass $M_T$ (Holsapple, et al., 2002). $Q_s$ is obtained by impact experiments. A few sample values of $\Gamma$ for $10^{-3}$ kg targets at impact speed of 20 km s$^{-1}$ are $\Gamma = 2.2\ 10^5$ for crystalline rock (Hörz et al., 1975) and $9\ 10^5$ for gypsum and pyrophyllite (Nakamura et al., 2015). For porous target bodies Love et al. (1993) found that $\Gamma \sim (1\text{-}f)^{3.6}$, where $f$ is the porosity of the target material.

For example a porous aggregate particle of 50% porosity (half the density of the solid building block material) has a factor ~10 smaller $\Gamma$ value than the solid material.

### 4.2. Low speed collisions

An important step of dust evolution from its formation in the atmospheres of cool red giant stars to meteorites and interplanetary dust is the accretion of interstellar dust in the early solar system. For several decades important experimental studies have been performed in order to better understand the build-up of planetary bodies from dust and gas during the formation of the planetary system (cf. Blum et al., 2000). Two major phases in the formation of solid planetary bodies can be distinguished: first, the aggregation of dust particles and clusters in the solar nebula caused by low-velocity mutual collisions to kilometre-sized planetesimals and second, the accretion of planetary cores and planets due to intermediate and high-velocity planetesimal collisions and gravitational attraction. A comprehensive review of the dust growth in protoplanetary discs is given in Blum (2019). Here we only want to stress the importance of laboratory investigations for the first step of accumulation of interstellar dust to larger aggregates.

Laboratory studies were performed to empirically determine sticking and fragmentation efficiencies in low velocity collisions of single dust grains and aggregates, and in intermediate to higher-velocity impact processes. The evolution of morphological structures of growing dust aggregates is an important aspect of these studies. It turned out that, after a period of rapid collisional growth of porous dust aggregates to sizes of a few centimetres, the protoplanetary dust particles are subject to bouncing collisions, in which their porosity is considerably decreased (Blum et al., 2010, 2019). According to these studies direct formation of kilometre-sized planetesimals by collisional sticking is unlikely, implying that collective effects, such as the streaming instability and the gravitational instability in dust-enhanced regions of the protoplanetary disc, are the best candidates for the processes leading to planetesimals. Studies of freshly released dust from comet 67P/Churyumov-Gerasimenko are underway to compare them with the predicted structures of aggregates during comet formation (Ellerbroek et al., 2017)

### 4.3. Environmental effects

Meteoroid bombardment of airless planetary surfaces comminutes the surface material and causes the build-up of a regolith layer. Regolith consists of impact-generated fragments which did not escape the gravity of the planetary body. Further impacts on to the regolith cause gardening and redistribution of material on the surface. Also UV and plasma exposure of regolith surfaces causes mobilization, transport, and ejection of dust particles from the surface. These processes are subject to extensive theoretical and laboratory studies (Szalay et al., 2019).

Electrostatic dust transport was first suggested five decades ago to explain the images of the Surveyor 5, 6 and 7 missions showing a lunar horizon glow: on the Moon a bright hovering cloud was observed shortly after sunset (Criswell, 1973; Rennilson and Criswell, 1974). Other examples showing evidence for dust transport across vast regions without winds or flowing water are the intermittently appearing radial spoke features first seen by the Voyagers above the rings of Saturn (Smith et al., 1981; 1982), and the accumulation of fine dust in “ponds” on the surface of asteroid Eros imaged by the NEAR-Shoemaker mission (Robinson et al., 2001).

In all these examples, both dust charges and electric fields predicted from current surface-plasma interaction models cannot give an explanation. However, Wang et al. (2016) recently observed jumping dust particles of about 100 μm diameter above a dusty surface when it was exposed to UV radiation or to plasmas in vacuum in the laboratory (Fig. 16). These experiments explained that the interaction of a dusty surface with ultraviolet radiation and/or plasmas is a volume effect. The emission and re-absorption of photo- and/or secondary electrons from the walls of micro-cavities formed between insulating dust particles below the surface are responsible for generating unexpectedly large charges and intense electric fields. The experimenters observed particles jumping to several millimetres height and dust ejection speeds up to 0.6 m $s^{-1}$. In the low gravity environment of asteroids and comets this effect could lead to dust jumps of several 100 m height and similar distances. The results indicate that electrostatic dust transport may be efficient in shaping the surfaces of airless bodies, such as surface morphology and porosity, and it may lead to space weathering.

Space weathering is another regolith process, which describes the darkening and reddening of airless planetary surface materials with time, together with changes to the depths of absorption bands in their optical spectra. This process has been invoked to explain the mismatched optical spectra of lunar rocks and regolith, and between those of asteroids and meteorites. The formation of nanophase iron particles on regolith grains as a result of micrometeorite impacts or irradiation by the solar wind has been proposed as the main cause of the change in the optical properties. Simulations of dust impacts by nano-second-pulse laser irradiation of olivine grains reproduced the optical changes. By observations with a transmission electron microscope Sasaki et al., (2001) found within the vapor-deposited rims of olivine grains nano-phase iron particles similar to those observed in the rims of space-weathered lunar regolith grains. The Hayabusa 2 sample return mission arrived at the carbonaceous asteroid Ryugu in 2018. Images revealed a rough asteroid surface with abundant boulders which unexpectedly lacks fine-grained regolith (Watanabe et al, 2019; Sugita et al. 2019, Jaumann et al, 2019). The spacecraft successfully deployed three landers to the asteroid surface and collected samples from two sites which will be returned to Earth in December 2020. The OSIRIS-REx sample return mission is currently orbiting the asteroid Bennu and mapping at centimetre resolution its surface with cameras, visible, infrared, and X-ray spectrometer (Enos and Lauretta 2019). In July 2020 the spacecraft will take regolith samples from the surface and return them to Earth in 2023.

### 4.4. Dust release from dirty snow

Already at the time when Giotto reached comet Halley in 1986, ESA mentioned the planetary Cornerstone mission Comet Nucleus Sample Return (CNSR) in its Horizon 2000 plan. CNSR should have followed NASA's Comet Rendezvous Asteroid Flyby mission (CRAF) as the next step in comet research. A joint ESA-NASA study group was set up to define the science of CNSR and to identify steps necessary to technologically implement such a mission. At the same time some members of the science team proposed to the German Science Foundation (DFG) a project to study how an ice-dust mixture reacts to solar irradiation in vacuum and to use the big solar simulation chamber left over from the Helios mission at DLR Cologne (Grün, 1991). The comet simulation experiment (KOSI) was selected and became part of a special DFG project “Small bodies in the solar system”. About 50 experimental and theoretical scientists and students and 10 technicians from 8 science institutes in Germany and 12 participating scientists from Austria, France, Israel, Netherlands, SU/Russia, and USA took part in the project (Grün et al., 1992, Sears et al., 1999). Related experimental investigations took place in Tel Aviv (Bar Nun et al. 2003), Graz (Kömle et al. 1996), and Dushanbe (Ibadinov et al. 1991). Eleven KOSI experiments were performed from May 1987 to May 1993. During insolation of the dirty ice, gas and dust emissions were determined. The initially homogeneous loose ice dust mixture decomposed into a stratified composition. At the surface a hot coherent but very fluffy dust mantle (~100 $kg/m^3$ density) developed over a solidified porous water ice layer (Stöffler et al. 1991). Below the hard ice layer more volatile ices like $CO_2$ ice were found which overlaid the cold original sample material. The emitted dust particles were mostly fragments of the dust mantle consisting of aggregates of the original building blocks mixed into the ices (Fig. 17, Kochan et al., 1990, 1998, Grün et al., 1993a). These particles resembled the fluffy IDPs collected in the Earth’s stratosphere (cf. Fig. 8). Recently, Rosetta observations stimulated a renewed interest in the results from laboratory studies of cometary processes (Gudipati et al., 2015, Poch et al., 2016, Jost et al., 2016).

# 5. Understanding the Zodiacal Cloud

Understanding the meteoroid environment is the subject of dust modelling. The models describe the life of meteoroids from their formation to their disappearance. Scientists already tried to understand the observed phenomena in the era when only ground based optical observations of dust and meteoroids were available. Foremost, Fred Whipple influenced the dust field with his ground-breaking work on comets (Whipple, 1950), their relationship to meteors (Whipple, 1951), and through his studies of the meteoritic complex as a whole (Whipple, 1967).

Understanding the dynamics of meteoroids is an important aspect of any modelling the meteoroid environment. Based on the fact that dust in space is subject to gravitational interaction, Öpik described the gravitational scattering of an interplanetary object by a planet (Öpik, 1951). While his work aimed at the distribution of interplanetary matter, it became also the basis of interplanetary travel to other planets using gravity assist. Fig. 18 shows the eccentricity vs. semi-major axis diagram of the small body reservoirs (Kuiper belt objects and asteroids) and their relation to the planetary scattering zones. Jupiter family comets are related to Trans-Neptunian objects via the Centaurs.

In addition to gravity, small meteoroids are affected by radiation pressure (cf. Burns et al., 1979). Absorbing particles with sizes comparable to the wavelength of visible sunlight (~0.5 μm) are affected most strongly, and the radiation pressure force may exceed solar gravity ($\beta > 1$). In this case particles leave the solar system on hyperbolic trajectories. Such beta meteoroids were first identified in the data from the Pioneer 8 and 9 space probes (Zook and Berg, 1975). The effect of radiation pressure was earlier applied to comet tails by Finson and Probstein (1968a, b). A comet tail can be described by an array of syndynes and synchrones where particles are sorted according to their repulsive force from solar radiation pressure and their ejection time from the nucleus, respectively. The secular effect of radiation pressure on meteoroid orbits, the Poynting-Robertson effect, was quantified by Wyatt and Whipple (1950). Through the Poynting-Robertson effect, interplanetary particles lose angular momentum and orbital energy and eventually spiral into the Sun. In most space environments, dust particles are electrically charged (Horanyi, 1996), and, hence, small, submicrometer-sized dust particles are strongly affected by their interaction with the interplanetary magnetic field (e.g. Morfill and Grün, 1979a, b).

Finally, the life times of meteoroids are limited by mutual collisions. Dohnanyi (1970) demonstrated that catastrophic collisions among meteoroids (i.e. the impact of a small meteoroid completely shatters the bigger meteoroid) are more effective in limiting the life time of sporadic meteoroids than erosive collisions (i.e. the impact of a small meteoroid generates just a crater on the bigger meteoroid). Dohnanyi (1969) also showed that the mass distribution of smaller asteroids (<100 km in size) is in a collisional quasi-steady-state: in any given size interval the number of destroyed asteroids equals the number of fragments generated by collisions from bigger asteroids. In this case the cumulative mass distribution has the form $m^{-\alpha}$ with $\alpha \approx 5/6$. The steeper mass distribution ($\alpha > 5/6$) of sporadic meteoroids is unstable and changes with time unless there is a sufficient source for such particles.

The first quantitative empirical dust model was deveoped by Cour-Palais (1969), aimed at describing to space engineers the meteoroid environment in near-Earth space. It was based on results from the Harvard Radio Meteor Project (summarized by Southword and Sekanina, 1973) and from the penetration detectors on board the Explorer 16, 23, and the Pegasus satellites. The model provided the flux of meteoroids in interplanetary space and in the near-Earth environment, taking into account gravitational enhancement and Earth shielding effects. Grün et al. (1985) published an interplanetary flux model on the basis of lunar micro-crater statistics and previous satellite data. It covered meteoroid masses from $10^{-18}$ to 1 g and assumed an average collision speed of 20 km $s^{-1}$ at 1 AU. It was demonstrated that the life time of meteoroids with masses below $10^{-9}$ kg (~100 μm) is dominated by the Poynting-Robertson effect whereas bigger particles are destroyed by mutual collisions. Results from impact experiments into basalt by Fujiwara et al. (1977) were used to quantify the collisional destruction of meteoroids. At 1 AU the collisional life time of centimetre-sized meteoroids was calculated to be approximately $10^4$ years.

In 1993 Divine published his “Five populations of interplanetary meteoroids” model (Divine, 1993). It was an empirical model based on data from radar meteor surveys, the previous interplanetary flux model by Grün et al. (1985), and space probe impact detector measurements ranging from 0.3 AU (Helios) to 20 AU (Pioneer 10). The model consistently described the interplanetary micrometeoroid environment via five distinct dust populations defined by their separable distributions in mass and orbital elements. Meteoroid fluxes, densities, and directional properties were calculated throughout the solar system using Keplerian dynamics. This model was later updated by Staubach et al. (1997) in order to include radiation pressure effects and thereby better represent the small particle flux measurements by Ulysses, Galileo, and other in situ dust measurements. A comprehensive review of the earlier dust models was given by Staubach et al. (2001).

NASA's Meteoroid Engineering Model, MEM, was developed by Jones (2004) and McNamara et al. (2004) on the basis of sporadic meteor observations by the Canadian Meteor Orbit Radar (CMOR, Jones and Brown, 1993) together with zodiacal light observations from Helios and lunar crater statistics. The model used orbital element distributions from meteor populations that were tied to actual comet families and asteroids (Jones et al., 2001). Since meteors must intersect the Earth orbit, this model can only be used to predict fluxes, speeds, and directions from approximately 0.5 to 2.0 AU.

A first evolutionary model of the sporadic meteoroid background inside Jupiter's orbit was published by Dikarev et al. (2004) and was implemented in ESA's Interplanetary Meteoroid Environment Model, IMEM. Contrary to earlier attempts, this model starts from the orbital elements of the dominant known sources of interplanetary dust: Jupiter family comets and asteroids. With this model the era of descriptions of the interplanetary dust cloud by just a handful of dust populations is left and only small computer programs can evaluate the dust densities and fluxes at any desired position in space. Six populations of dust from asteroids, 72 populations of dust from comets and other parent objects on Jupiter-crossing orbits, and one stream of interstellar dust, were introduced in the IMEM model (for a full description see Dikarev et al., 2019). The model assumes that big meteoroids ($\geq 10^{-5}$ g) stay in the orbits of their parent objects (because collisions destroy them before they are removed by dynamical effects) while the orbits of smaller meteoroids evolve under planetary gravity and the Poynting-Robertson effect. Thermal radiation measurements by the COBE DIRBE instrument (Kelsall et al., 1998), in situ data from the dust instruments onboard Galileo and Ulysses (Grün et al., 1997), and lunar microcrater distributions were used to calibrate the contributions from the known sources. However, due to an inconsistency between the COBE DIRBE observations and the Galileo and Ulysses in situ detections, the Galileo and Ulysses data were adjusted such that the grain sizes were increased by a factor of 2.5 with respect to the usual instrument calibration (Dikarev et al., 2019). Attempts to include meteor orbits from the Advanced Meteor Orbit Radar AMOR in the model (Galligan and Baggaley, 2004) failed since the AMOR orbital distributions were incompatible with the COBE latitudinal IR brightness profile under the assumptions of the model.

Recently, Nesvorny et al. (2010, 2011a, 2011b) published a dynamical model of the zodiacal cloud and sporadic meteors. The sources of this evolutionary model are asteroids, Jupiter-family comets, Halley-type comets, and Oort-cloud comets. The model describes zodiacal infrared observations by the Infrared Astronomical Satellite (IRAS, Hauser et al., 1984), distributions of radar meteors from the Canadian Meteor Orbit Radar (CMOR, Campbell-Brown, 2008), and properties of micrometeorites recovered in Antarctica (Engrand and Maurette, 1998). The authors found that Jupiter family comets are the main source of meteor concentrations arriving at the Earth from the helion and antihelion directions (cf. Fig. 6), as well as the sources of micrometeorites in our collections (micrometeorites collected in Antarctic ice have primitive carbonaceous composition). At Earth, asteroidal dust contributes less than 10%. However, the authors rather crudely extrapolated from the collision life time given by Grün et al. (1985) and applied a fitting parameter which multiplies the collision life times. They found that the orbits of some particle populations must be much further evolved by the Poynting-Robertson effect than the original collisional life times would allow. The fitting parameter must be 10 to 30 in order to be compatible with infrared latitudinal brightness profiles and to match the speed distribution measured by the radar meteor observatories. They even required that about 1% of particles from Oort cloud comets orbitally evolve by Poynting–Robertson drag to reach orbits with semi-major axis a ~1 AU. Such meteoroids are expected to produce meteors with radiants near the apex of Earth's orbital motion. Nesvorny et al. (2011b) also include a set of Jupiter family and Oort cloud comets that are similar, but dynamically evolved beyond their normal activity life times. This is to account for a dust component resulting from the spontaneous disruption of such dormant comets. This model has recently been applied in the inner solar system to develop a comprehensive model of the meteoroid environment around Mercury and to explain structural and temporal variability in Mercury's exosphere (Pokorny et al., 2018).

Motivated by the need to understand the meteoroid hazard to human exploration beyond Earth vicinity, ESA supported the IMEX study (Interplanetary Meteoroid Environment for eXploration) at the Institut für Raumfahrtsysteme in Stuttgart (Soja et al., 2014, 2015a,b). The study was based on the observations by Reach et al. (2007) with the Spitzer Space Telescope. The latter authors found that 30 out of 34 Jupiter-family comets had known debris trails. Thus, the detection rate exceeded 80%, indicating that debris trails are a generic feature of short-period comets. In the IMEX study the evolution of dust streams from 362 Jupiter family comets, 40 Halley-type comets and 18 Encke-type comets were calculated between previous five apparitions and the year 2080. A test of the model with the Leonid meteor storm demonstrated that it predicts within 20 minutes the maximum and the duration of the November 1999 storm. The amplitude of the profile, however, was not matched well because of uncertainties of the cometary dust production rate and mass distribution. The application to the dust streams of comet 67P/Churyumov-Gerasimenko results in the prediction of the ejection velocities, the dust size index, and a dust production rate of this comet. Also a meteor stream from this comet is predicted at Mars for the year 2062. The meteoroid stream environment of the Earth in September 2049 is shown in Fig. 19.

Currently, the space agencies ESA and NASA are in the process of updating their interplanetary meteoroid models. The ESA Interplanetary Meteoroid Environment Model, IMEM2, has been developed at the Institut für Raumfahrtsysteme in Stuttgart (Soja et al., 2019). The orbit evolution of dust emitted from various types of sources (Jupiter family, and Halley type comets and asteroids) is followed for a million years and the collision history is determined (Soja et al., 2016). The model is fitted to the lunar micro-crater distribution (Morrison and Zinner, 1977, Morrison and Clanton, 1979), and meteor observational data (Campbell-Brown, 2008, Galligan and Baggaley, 2005). Calibration of dust densities and fluxes is provided by a fit to the absolute infrared brightnesses observed by COBE (Kelsall et al., 1998). In this model it was found that the collisional life times are significantly extended as compared to the ones assumed by Grün et al. (1985) for rocky meteoroid material. It was also determined that about 80% of the dust observable by infrared methods originates from Jupiter family comets, 20% from asteroids, and only 0.5% from Halley type comets. Both findings support the hypothesis that interplanetary dust particles have mostly a porous fluffy structure just like cometary particles have (cf. Figs. 1 and Fig. 5).

In the outer solar system, beyond Jupiter's orbit, the sources of dust are different, including a dominant Kuiper Belt component, but the particles are subject to the same perturbing and modifying effects as in the inner solar system.

An understanding of the dust cloud in this region is important for quantifying the interplanetary dust input in the Jovian and Saturnian systems: A comprehensive model has been built by Poppe, 2016 (see also Szalay et al., 2019).

In addition to the flow of interplanetary matter mostly released from comets and asteroids, there is a flow of interstellar grains through the planetary system. Following the earlier predictions (Bertaux and Blamont, 1976, Gustafson and Misconi, 1979, Morfill and Grün, 1979b) of the flow of electrically charged interstellar dust through the planetary system such flow was successfully detected by the Ulysses dust instrument (Grün et al., 1994). The models by Landgraf (2000), Sterken et al. (2012, 2013, 2015) and Strub et al. (2019) show strong spatial and temporal variations of the interstellar dust density due to its interactions with solar radiation pressure and the interplanetary magnetic field (Sterken et al., 2019).

The dynamical dust processes in circumplanetary environments of the outer planets are even more complex than in interplanetary space because of the proximity to the central body, the multitude of satellites some of which are active dust/ice emitters, the existence of rings, and the strong magnetospheric effects. Nevertheless, a detailed understanding of most of the observed phenomena has been developed. For a comprehensive review see Spahn et al. (2019).

Despite the great advances made in the last years in understanding the heliospheric dust environment, there remain many important questions to be answered. The overall goal is to collect a compositional dust inventory for the whole planetary system from the solar F-corona to the Kuiper belt and to characterize its interrelation with the planetary, interplanetary, and interstellar environment.

We have summarized the long history of dust research starting from the perception by many scientists that dust is dangerous to encounter, dirty and impure in its properties and difficult to quantify it individually and collectively. In our present understanding dust is a treasure chest of information on its sources and the processes that shape its properties throughout space and time. This book gives testimony that dust bridges astrophysics, planetary science, and life science.

# Tables

Table 1. In situ dust detectors in interplanetary space: distance of operation, mass sensitivity and sensitive area.

| Spacecraft | Launch | Distances | Mass threshold | Area |
|---|---|---|---|---|
| | Year | (AU) | (g) | ($m^2$) |
| Pioneer 8 | 1967 | 0.97 - 1.09 | $2x10^{-13}$ | 0.0094 |
| Pioneer 9 | 1968 | 0.75 - 0.99 | $2x10^{-13}$ | 0.0074 |
| HEOS 2 | 1972 | 1 | $2x10^{-16}$ | 0.01 |
| Pioneer 10 | 1972 | 1 - 18 | $2x10^{-9}$ | 0.26 |
| Pioneer 11 | 1973 | 1 - 10 | $10^{-8}$ | 0.26 |
| Helios 1/2 | 1974/76 | 0.3 - 1 | $10^{-14}$ | 0.012 |
| Galileo | 1989 | 0.7 - 5.3 | $10^{-15}$ | 0.1 |
| Hiten | 1990 | 1 | $10^{-15}$ | 0.01 |
| Ulysses | 1990 | 1 - 5.4 | $10^{-15}$ | 0.1 |
| GORID | 1996 | 1 | $10^{-15}$ | 0.1 |
| Cassini CDA | 1997 | 0.7 - 10 | $2x10^{-16}$ | 0.09 |
| Cassini HRD | 1997 | 0.7 - 10 | $3x10^{-13}$ | 0.006 |
| Nozomi | 1998 | 1 – 1.5 | $10^{-15}$ | 0.01 |
| New Horizons | 2006 | 2.6 - >45[*0] | $2x10^{-12}$ | 0.1 |
| IKAROS-ALADDIN[*1] | 2010 | 0.72 - 1.1 | $10^{-9}$ | 0.54 |
| LADEE[*2] | 2013 | 1 | $2x10^{-16}$ | 0.01 |
| BepiColombo MMO-MDC[*3] | 2018 | 0.31-0.47 | $1x10^{13}$ | 0.006 |
| DESTINY PLUS-DDA[*4] | ~2023 | 0.75 - 1.0 | $10^{-16}$ | 0.035 |
| MMX CMDM[*5] | ~2024 | 1 - 1.5 | ~$10^{-10}$ | ~1 |
| **Europa Clipper SUDA**[*6] | **~2025** | **5** | **$10^{-16}$** | **0.025** |

[*0] Jan. 2019

[*1] Interplanaetary Kite-craft Accelerated by Radiation of the Sun, Arrayed Large-Area Dust Detectors in Interplanetray Space

[*2] Lunar Atmosphere and Dust Environment Explorer

[*3] Mercury Magnetospheric Orbiter, Mercury Dust Monitor

[*4] Demonstration and Experiment of Space Technology for Interplanetary Voyage Phaethon Flyby and Dust Science Dust Analyzer

[*5] Martian Moons Exploration Circum-Martian Dust Monitor

[*6] Surface Dust Mass Analyzer

Table 2. Comet dust analyzers, mass sensitivity and sensitive area, and compositional resolution

| **Mission Instrument** | **Encounter** | **Comet** | **mass threshold** | **Area** | **Compositional resolution** |
|---|---|---|---|---|---|
| | **Year** | | **(g)** | **(m²)** | **M/ΔM** |
| **Giotto Dust Impact Detection System (DIDSY)** | **1986 1992** | **Halley, Grigg Skjellerup** | **$10^{-8}$** | **1** | **—** |
| **Giotto Particle Impact Analyzer (PIA)** | **1986** | **Halley** | **$2x10^{-15}$** | **0.0005** | **100** |
| **VeGa 1/2 Dust Counter and Mass Analyzer (DUCMA)** | **1986** | **Halley** | **$10^{-11}$** | **0.0075** | **—** |
| **VeGa 1/2 Particle Impact Mass Analyzer (PUMA)** | **1986** | **Halley** | **$2x10^{-15}$** | **0.0005** | **100** |
| **VeGa 1/2 Solid Particle Experiment (SP)-1** | **1986** | **Halley** | **$2x10^{-15}$** | **0.0081** | **—** |
| **VeGa 1/2 SP-2** | **1986** | **Halley** | **$10^{-11}$** | **0.05** | **—** |
| **Stardust Aerogel Collector** | **2004** | **Wild 2** | **$\sim10^{-12}$** | **0.1** | **>1000 (in laboratory)** |
| **Stardust Cometary and Interstellar Dust Analyzer (CIDA)** | **2004 2011** | **Wild 2 Tempel 1** | **$2x10^{-15}$** | **0.009** | **200** |
| **Stardust Dust Flux Monitor** | **2004 2011** | **Wild 2 Tempel 1** | **$10^{-5}$, $10^{-12}$ $10^{-8}$, $3x10^{-}$** | **0.7, 0.002** | **—** |
| **Instrument (DFMI)** | | | **12** | | |
| **Rosetta Grain Impact Analyzer and Dust Accumulator (GIADA)** | **2014-2016** | **Churyumov-Gerasimenko** | $10^{-7}$ | **0.01** | **—** |
| **Rosetta Cometary Secondary Ion Mass Analyzer (COSIMA)** | **2014-2016** | **Churyumov-Gerasimenko** | $10^{-7}$ | **0.0003** | **2000** |
| **Rosetta Micro-Imaging Dust Analysis System (MIDAS)** | **2014-2016** | **Churyumov-Gerasimenko** | $10^{-16}$ | $10^{-5}$ | **—** |
| **Rosetta/Philae Surface Electric Sounding and Acoustic Monitoring Experiment-Dust Impact Monitor (SESAME-DIM)** | **2014** | **Churyumov-Gerasimenko** | $10^{-4}$ | **0.007** | **—** |

# Figures

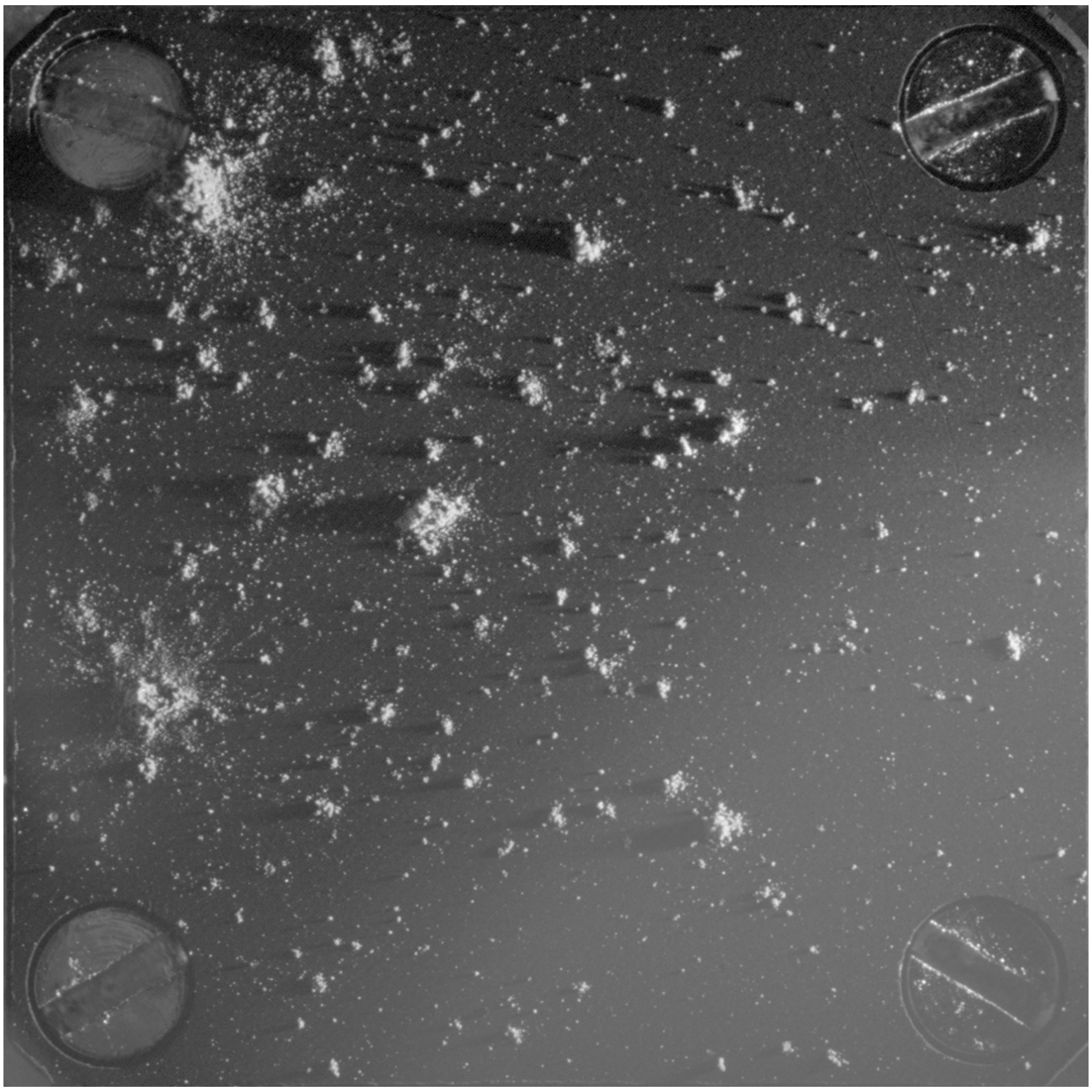

Figure 1. Comet particles collected and analyzed by the COSIMA instrument on board Rosetta. The particles were collected in between 10 April 2015 and 27 May 2015 at distances between 91 km and 321 km from the nucleus of comet 67P (at heliocentric distance between 1.89 and 1.55 AU), at a collection speed of ~10 m/s. For scale the screw head has 1.5 mm diameter (Langevin et al., 2016, Hilchenbach et al., 2016, ESA/ROSETTA/MPS FOR COSIMA TEAM MPS/CSNSM/UNIBW/TUORLA/IWF/IAS/ESA/BUW/MPE/LPC2E/ LCM/ FMI/UTU/LISA/UOFC/VH&S)

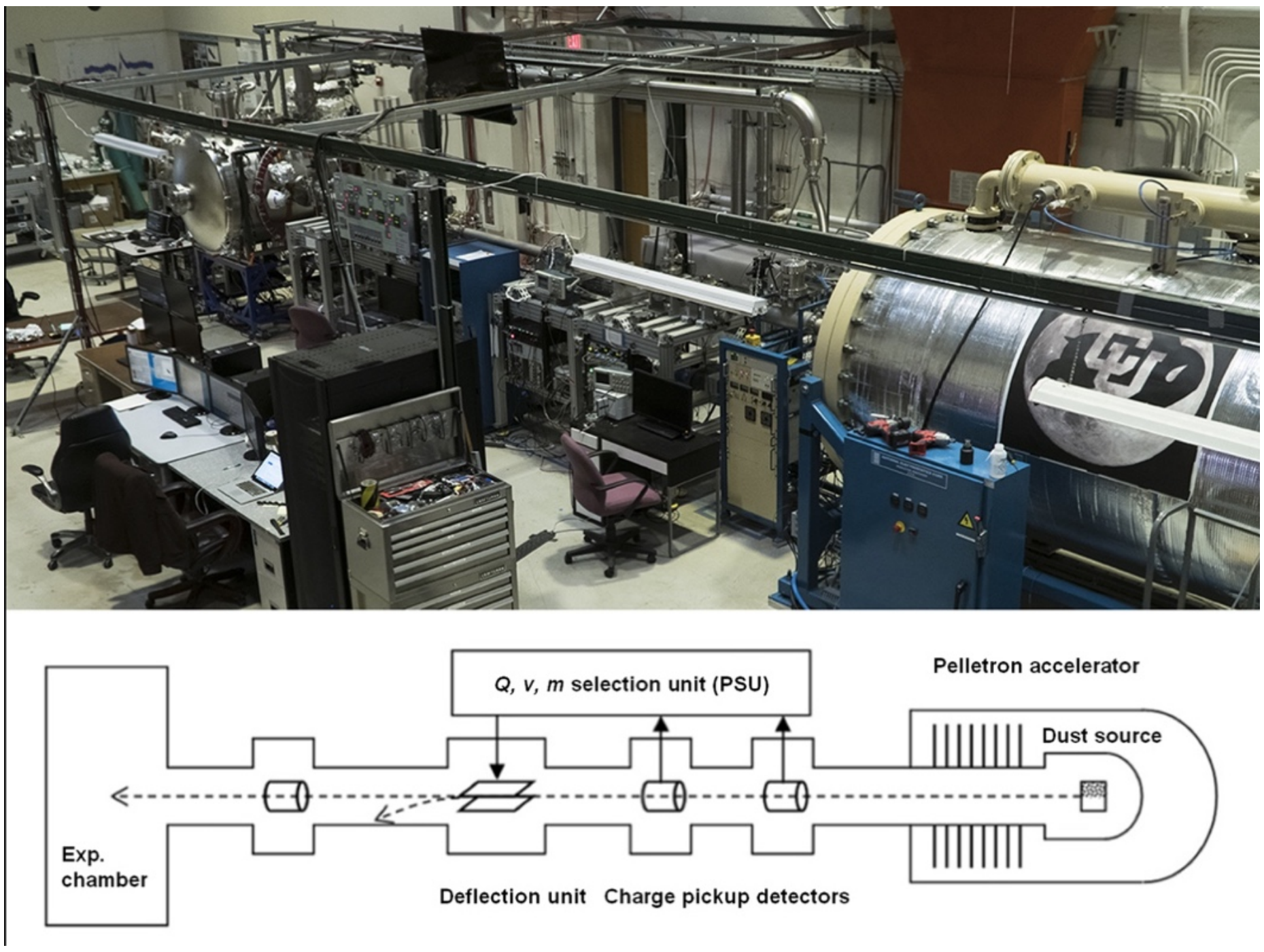

Figure 2. Dust accelerator at the University of Colorado/LASP with 3 MV acceleration voltage. The schematics are shown at the bottom (after Horanyi et al. 2016).

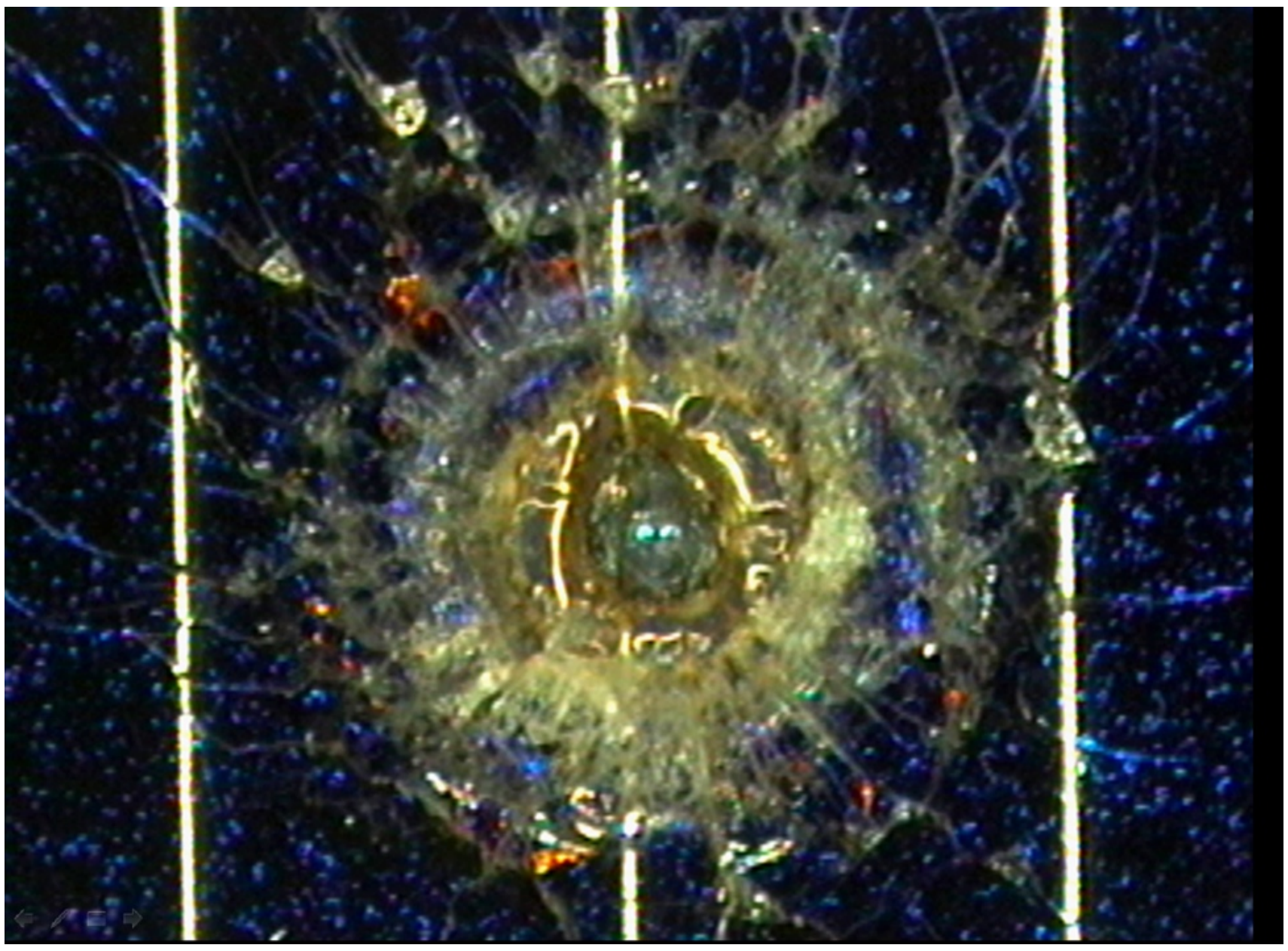

Figure 3. Space debris impact on to Hubble Space Telescope solar cells that were returned to ground (NASA image)

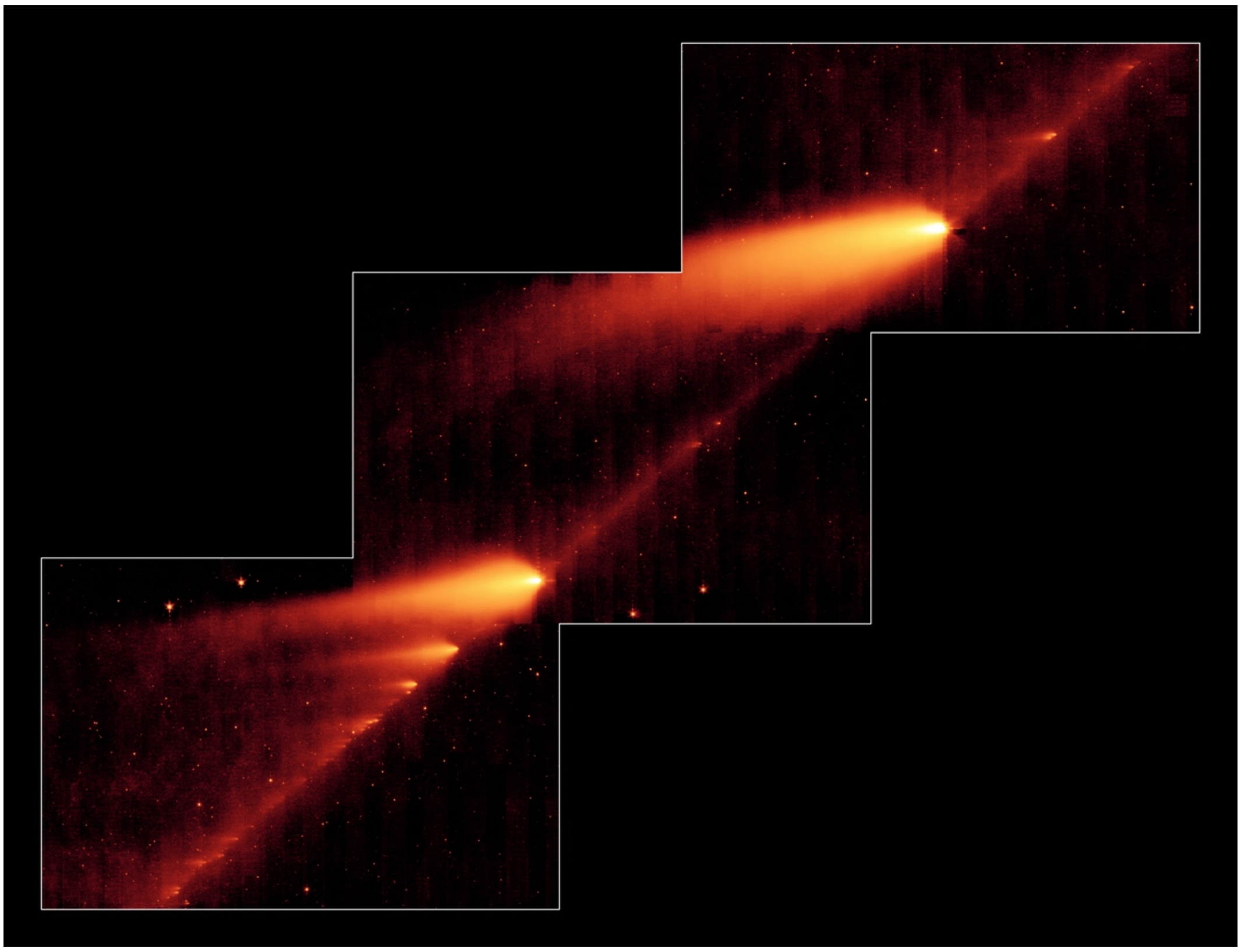

Figure 4. Breakup of comet 73P/Schwassmann-Wachmann (NASA Spitzer Space Telescope). Individual fragments are spread along the original orbit and display their own dusty tails (Reach et al., 2009). Comet 73P/Schwassman-Wachmann 3 images taken from May 4 to 6 show at least distinct 36 fragments. Credit: NASA/JPL-Caltech

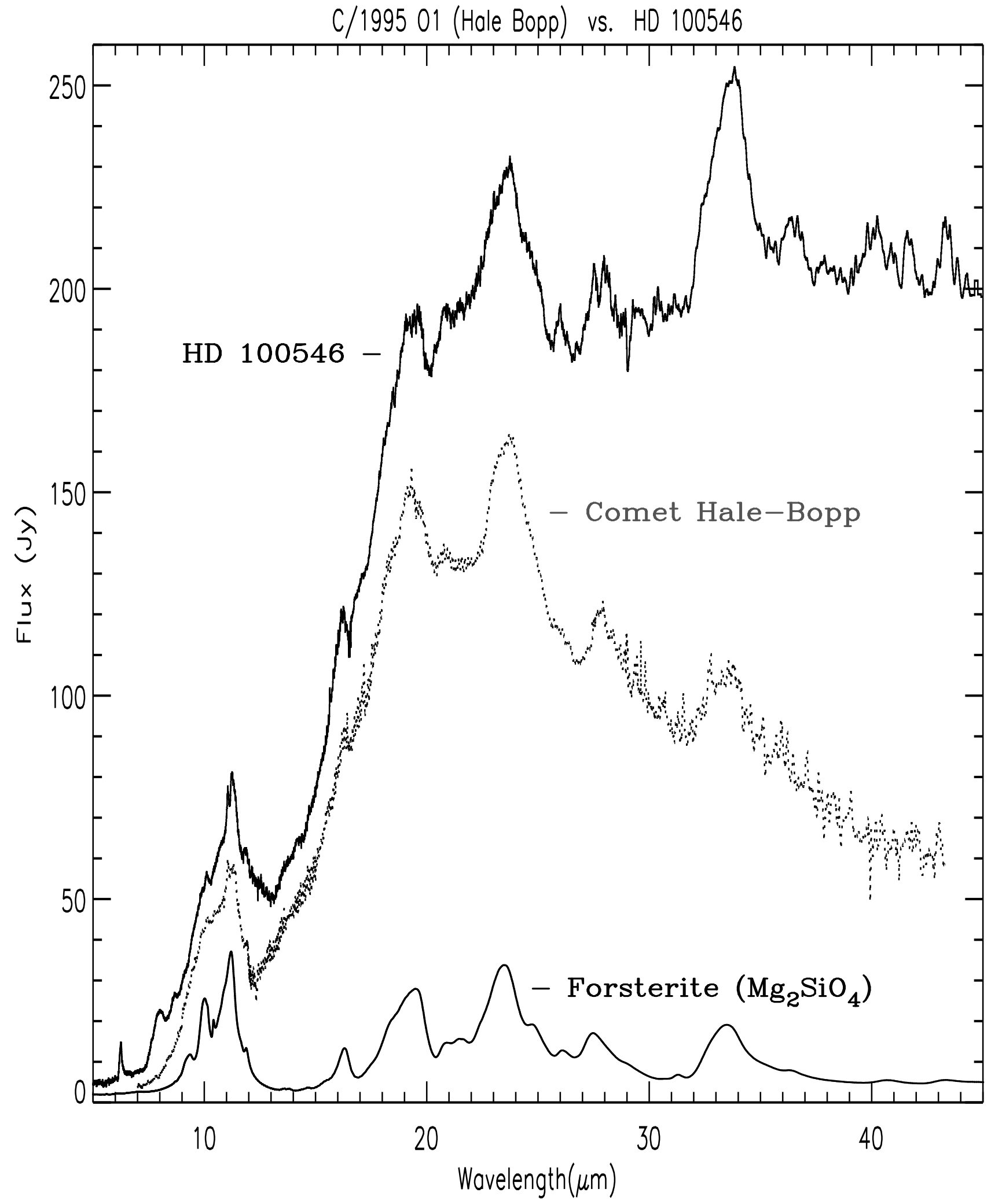

Figure 5. Infrared spectra of the Young Stellar Object HD 100546 (top curve) compared to those of comet Hale-Bopp (middle), both taken with the Infrared Space Observatory (ISO). The bottom curve shows a laboratory spectrum of forsterite [van den Ancker, 1999].

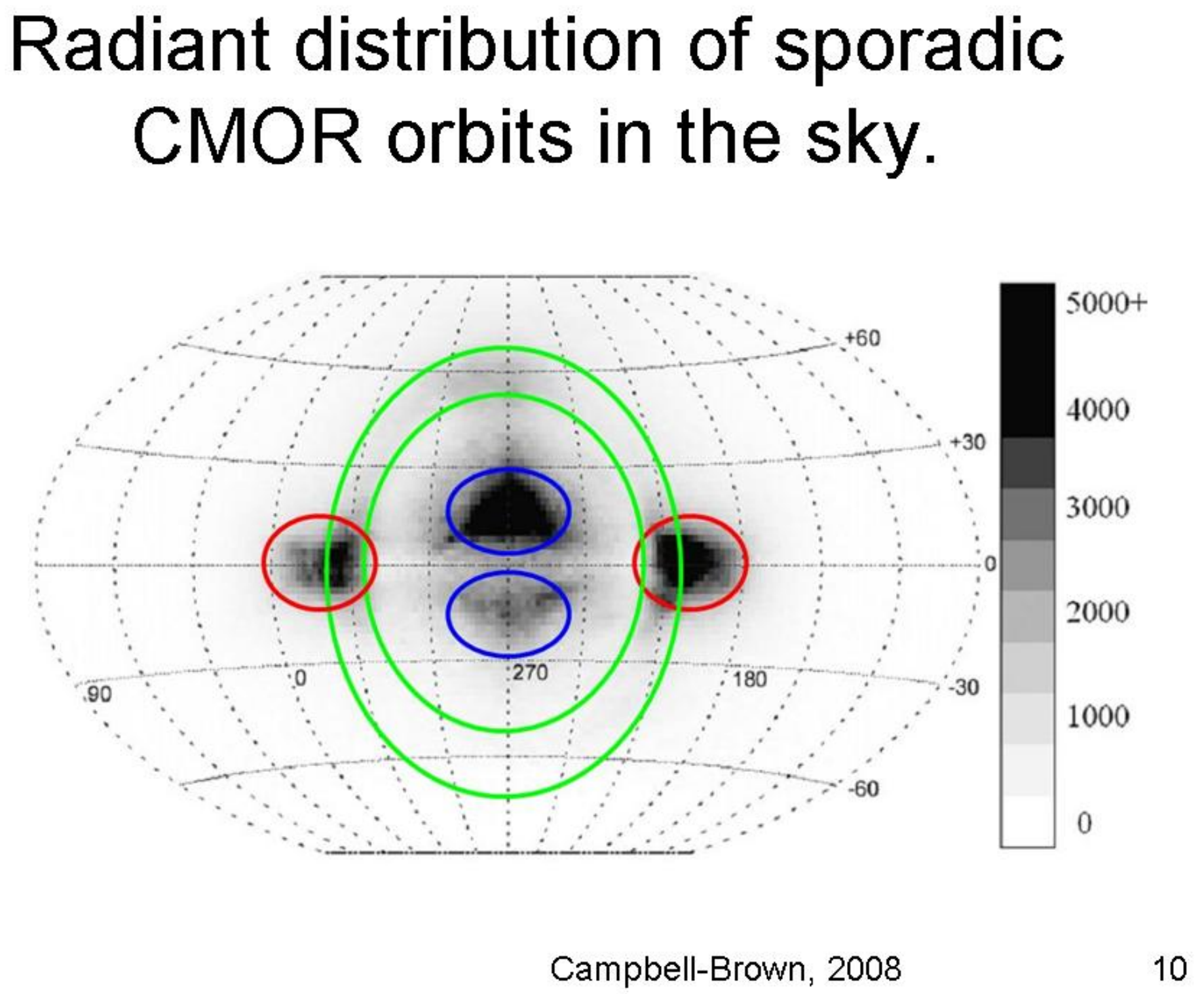

Figure 6. Apparent radiant distribution of sporadic CMOR orbits in the sky, corrected for in-atmosphere observing biases (after Campbell-Brown, 2008) with individual types of radiant concentrations indicated by colored ellipses. Red: helion (left) and antihelion (right), blue: north and south apex, and green: toroidal type. Because CMOR is located at 43° North latitude the southern components are reduced or not visible.

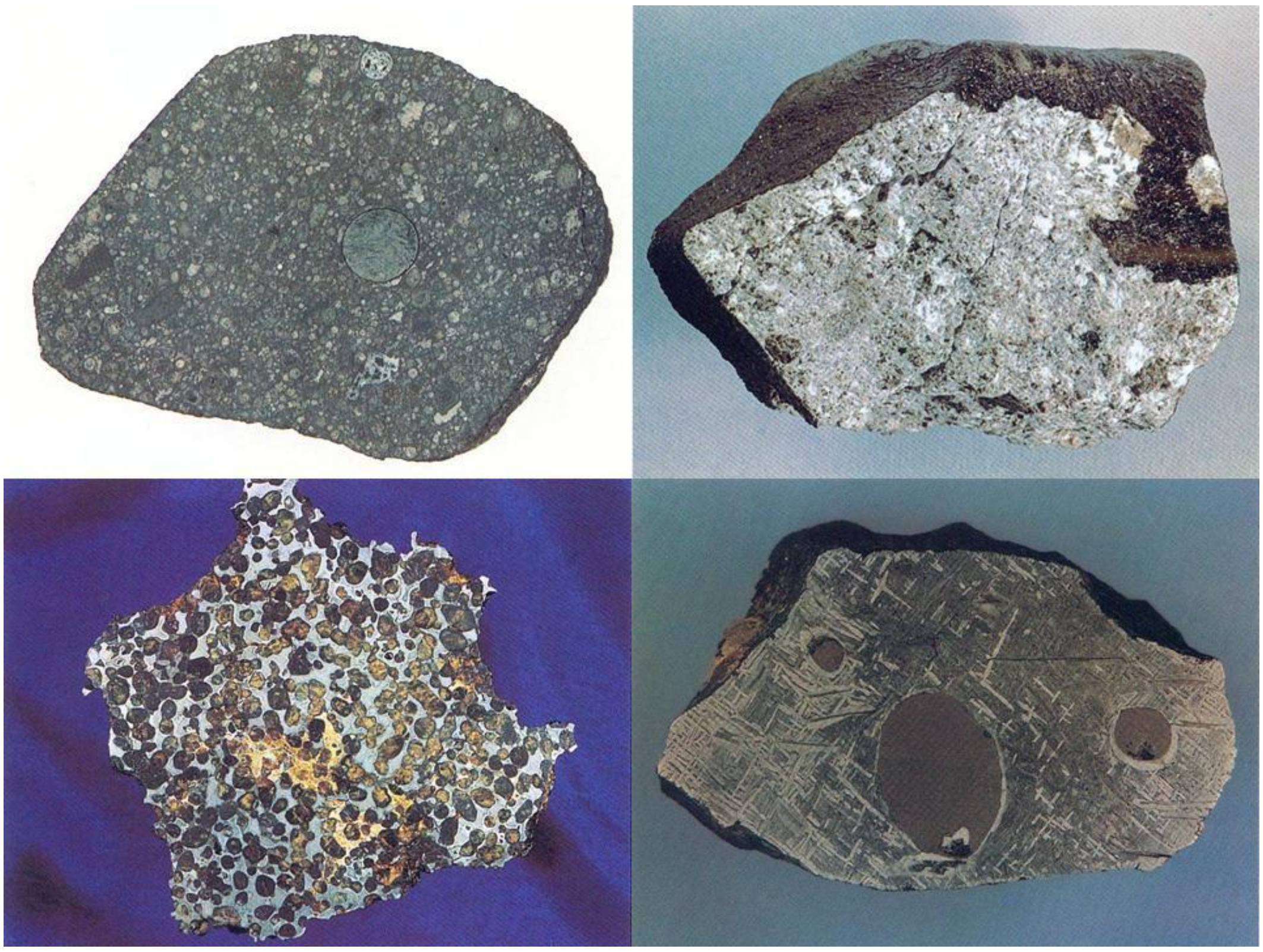

Figure 7. Meteorite types in clockwise order: carbonaceous chondrite, stony meteorite, stony-iron meteorites, iron meteorite (after Lipschutz and Schultz, Encyclopedia of the Solar System 2nd. Edition).

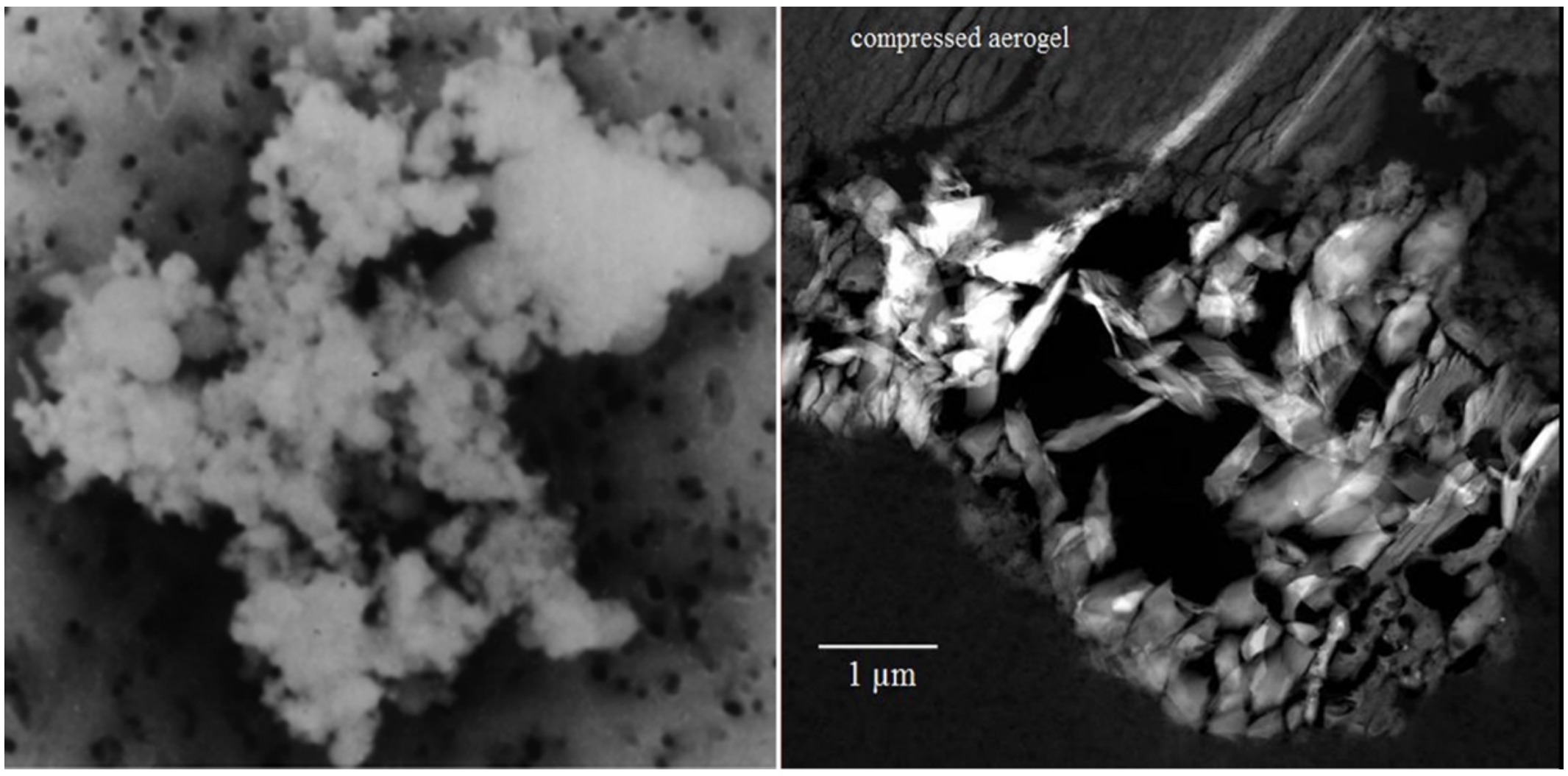

Figure 8. Chondritic dust particle collected in the Earth's stratosphere (left, NASA image). Single CAI particle collected from comet Wild 2 by NASA Stardust mission (right, cf. Zolensky et al. 2006)

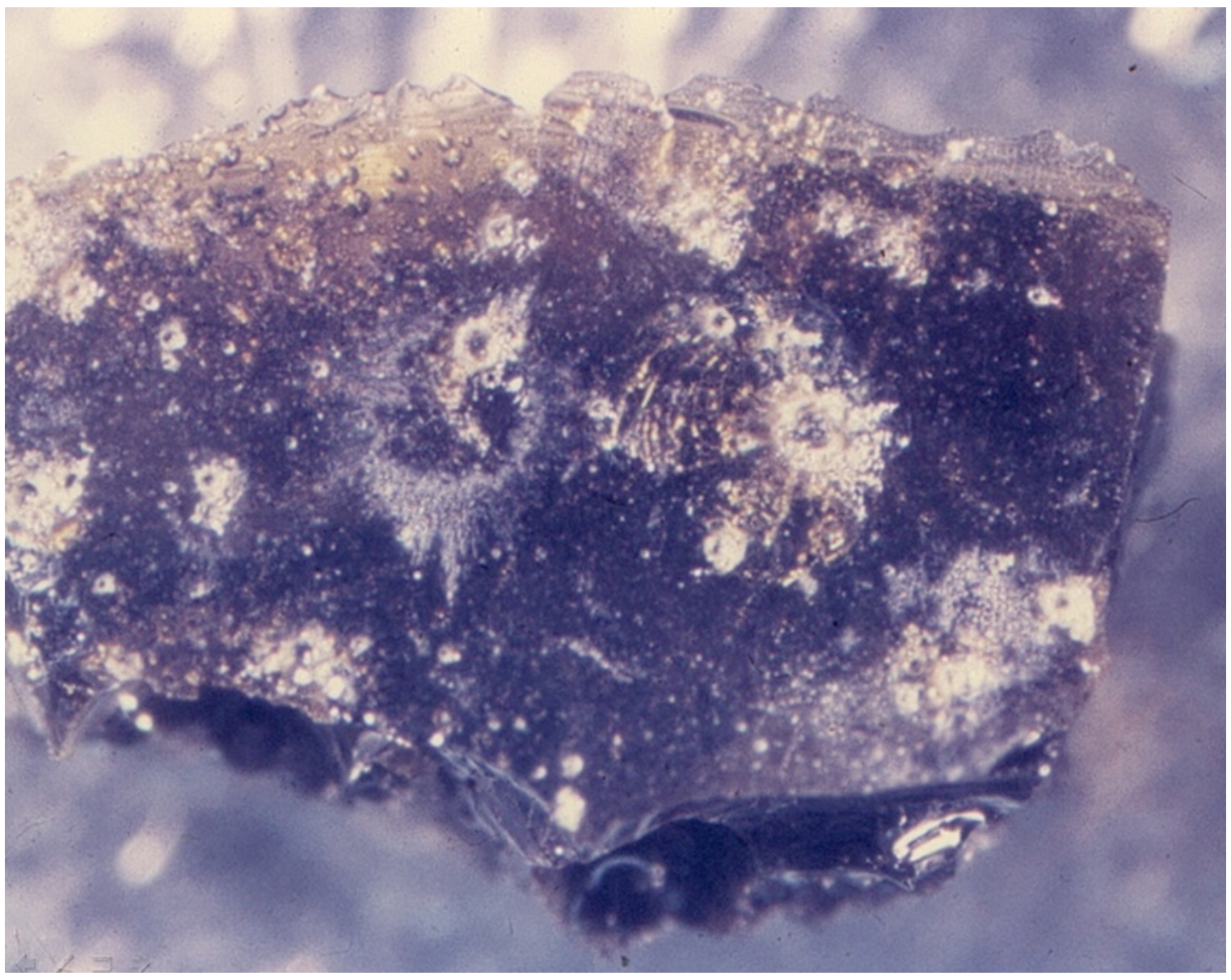

Figure 9. Microcraters on the glassy surface of a lunar rock sample (NASA image)

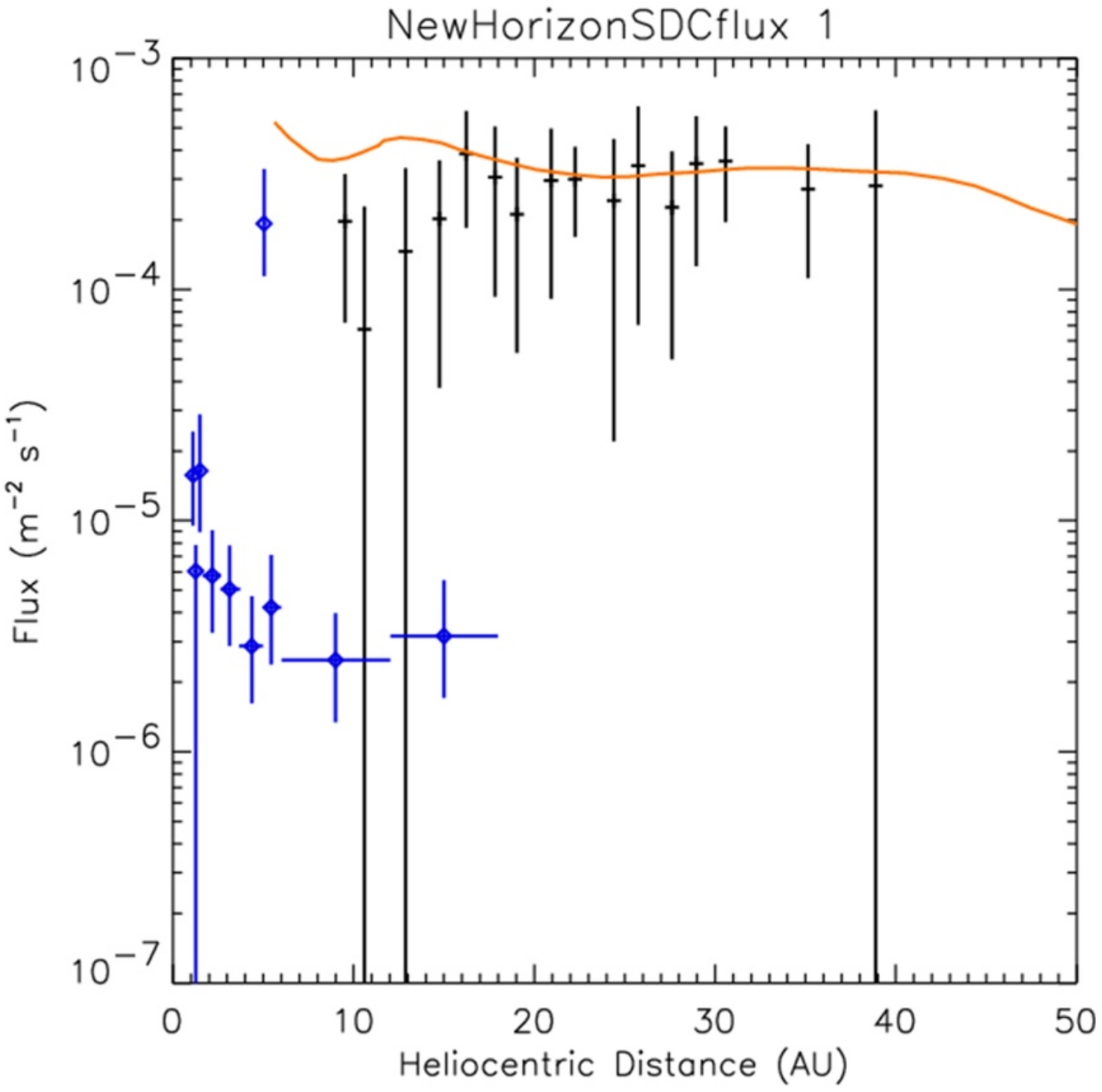

Fig. 10 Interplanetary dust flux in the outer planetary system. Black: flux onto New Horizons SDC for grains with radii >0.6 µm (Piquette et al. 2019). The red curves represents a model of the predicted SDC flux (Poppe, 2016). Blue: Pioneer 10 flux (Humes et al, 1980).

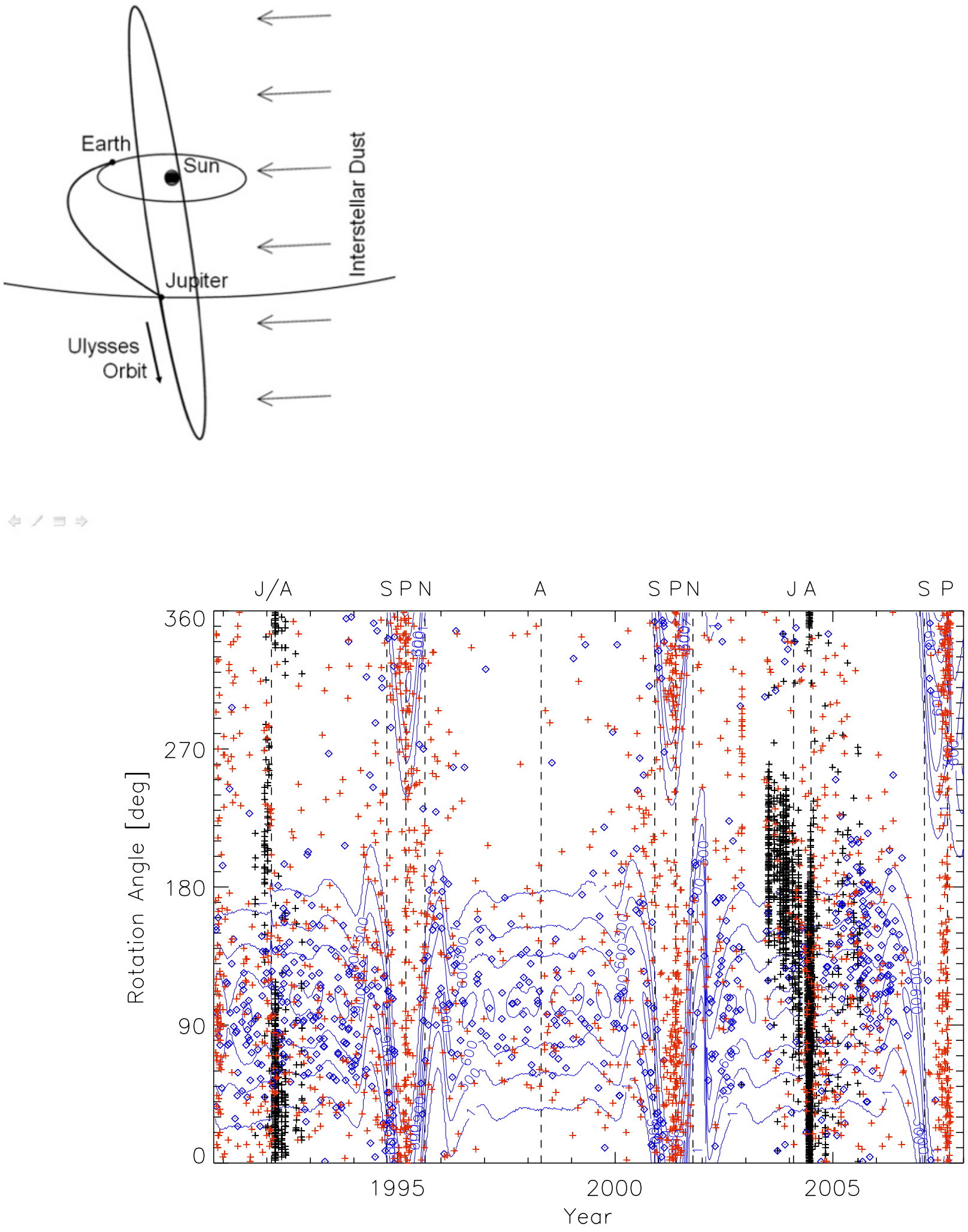

Figure 11. Left panel: Ulysses orbit. Positions of Earth at launch and of Jupiter at flyby. Interstellar dust flow is almost perpendicular to the Ulysses orbit plane. Right panel: Impact direction (rotation angle) verus time for all dust impacts detected by Ulysses between launch in October 1992 and the end of dust instrument operation in November 2007. Each symbol indicates an individual dust particle impact. Red crosses: Interplanetary dust; black crosses: Jupiter stream particles; blue diamonds: interstellar dust particles, blue contour lines show the effective sensor area for dust particles approaching from the upstream direction of interstellar helium. Vertical dashed lines and labels at the top indicate Ulysses's Jupiter flybys (J), perihelion passages (P), aphelion passages (A), south polar passes (S), and north polar passes of Ulysses (N) (adapted from Krüger et al., 2015a).

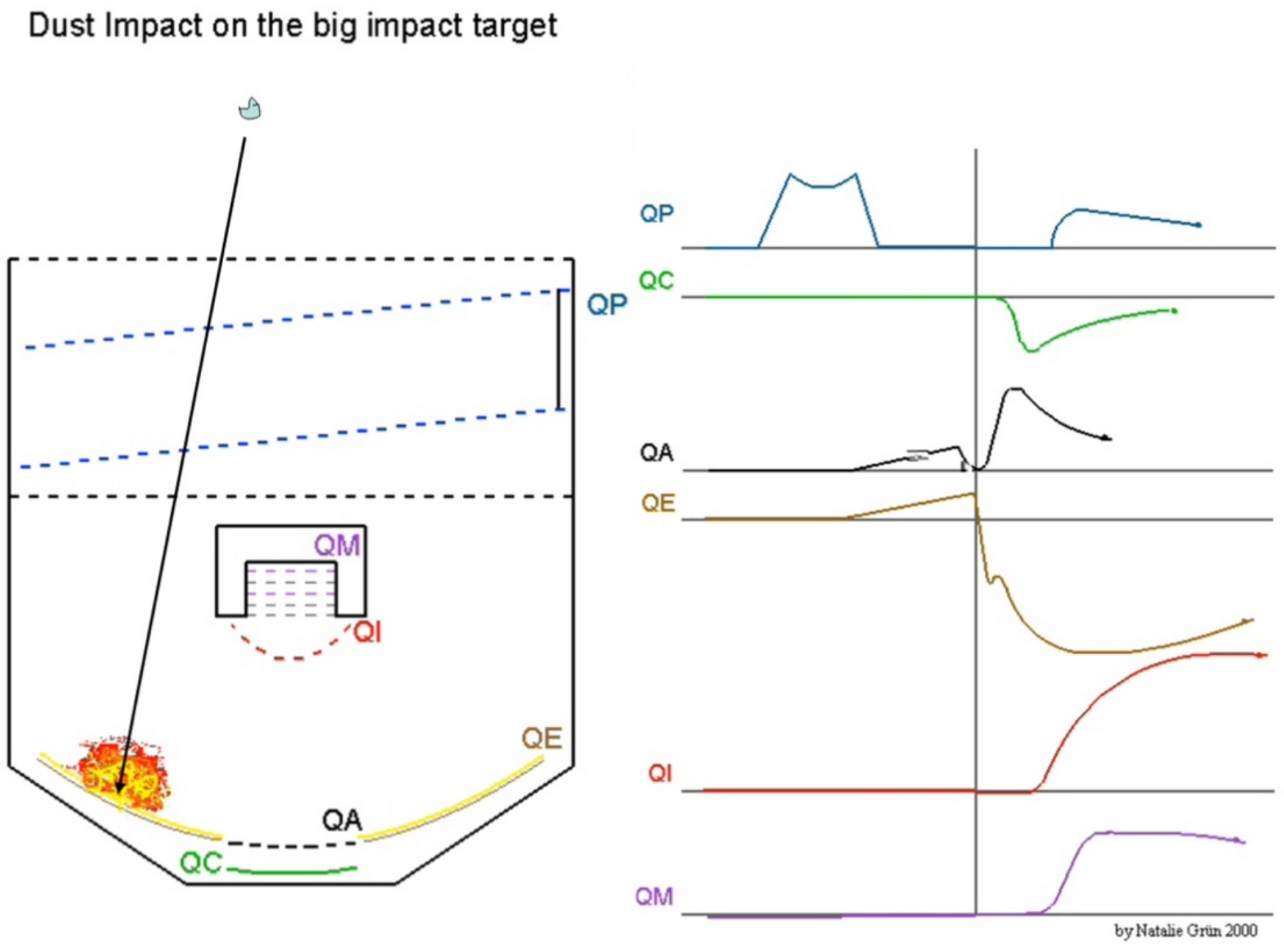

Figure 12. Cassini Cosmic Dust Analyzer, CDA, left. Instrument schematics with signals from an impact onto the large hemispherical target (right). Impacts onto the centre target will generate a mass spectrum of the generated ions (Srama et al., 2004).

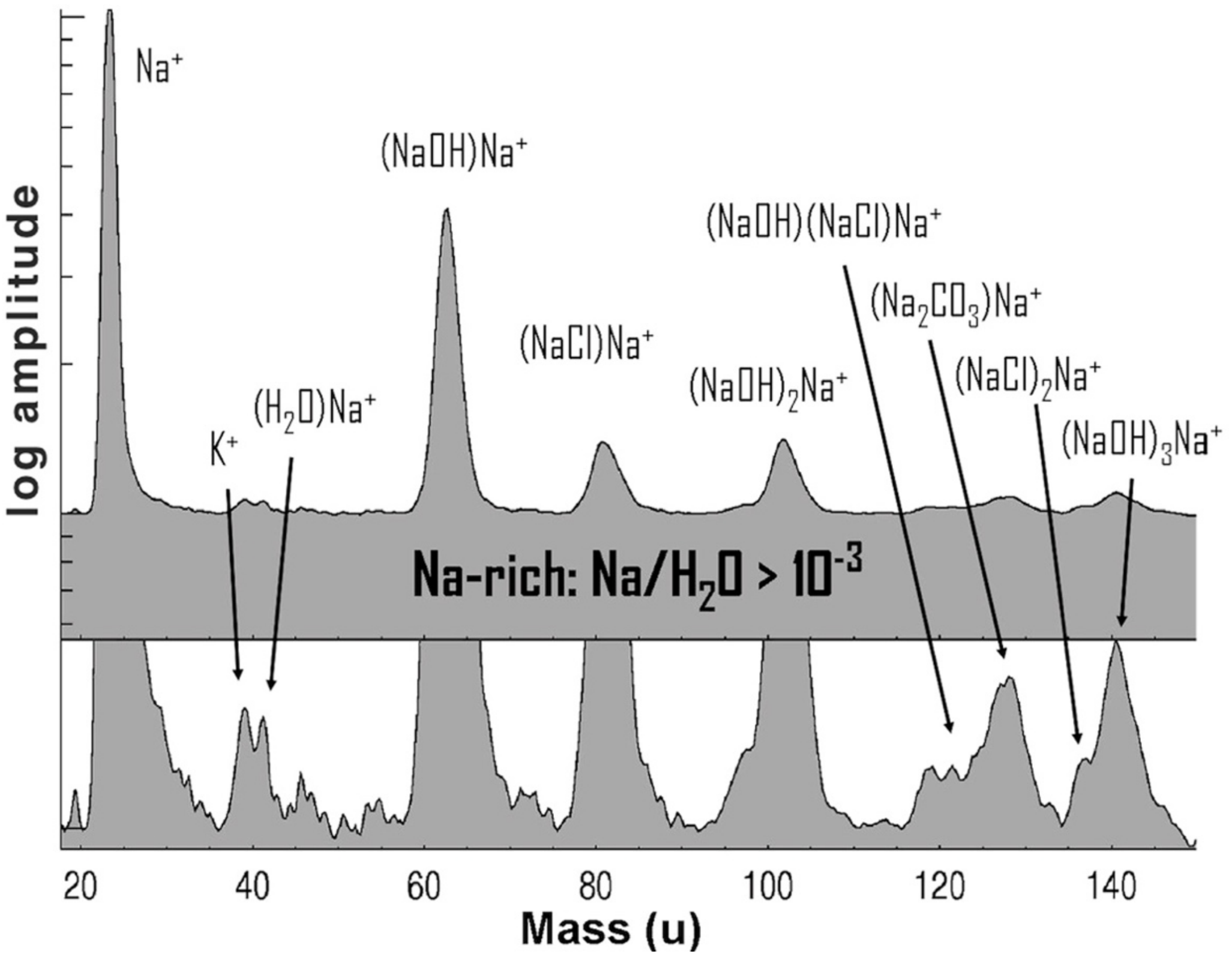

Figure 13. Co-added spectra of salt-rich water-ice particles measured by Cassini CDA in the E-ring of Saturn. Although the particles are still predominantly water, these spectra typically show very few pure water and Na-hydrate clusters, if any. They are characterized by an abundant Na+ mass line followed by a peak sequence of hydroxyl-cluster-ions (NaOH)n Na+ (cf. Postberg et al., 2009)

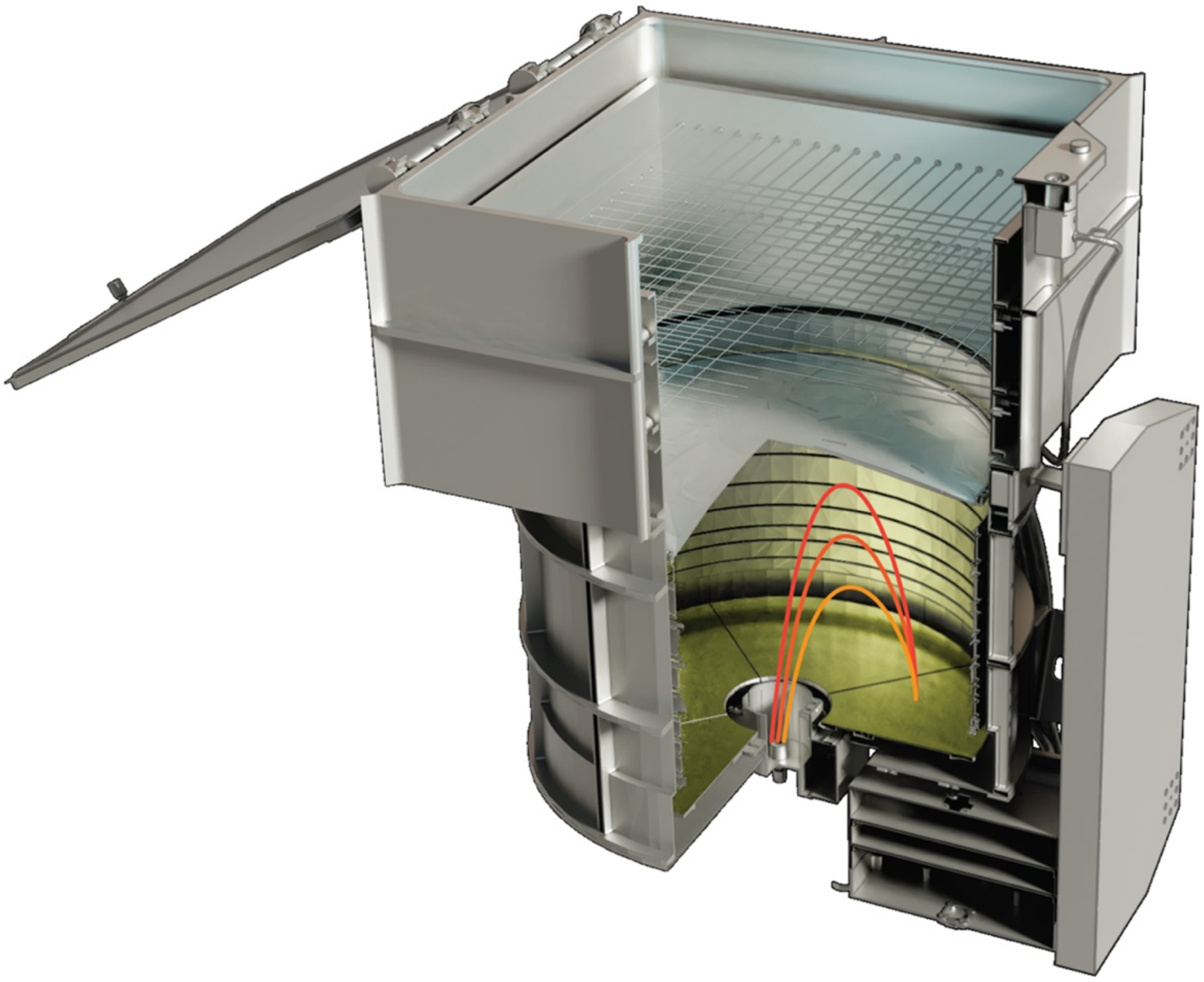

Figure 14. Dust Telescope consisting of a trajectory sensor (on top) in combination with a large-area mass analyzer (below). In the cut-out ion paths (red) from the impact on the target to the central ion collector are visible (from the FOSSIL proposal, Horanyi et al. courtesy Tibor Balint, JPL).

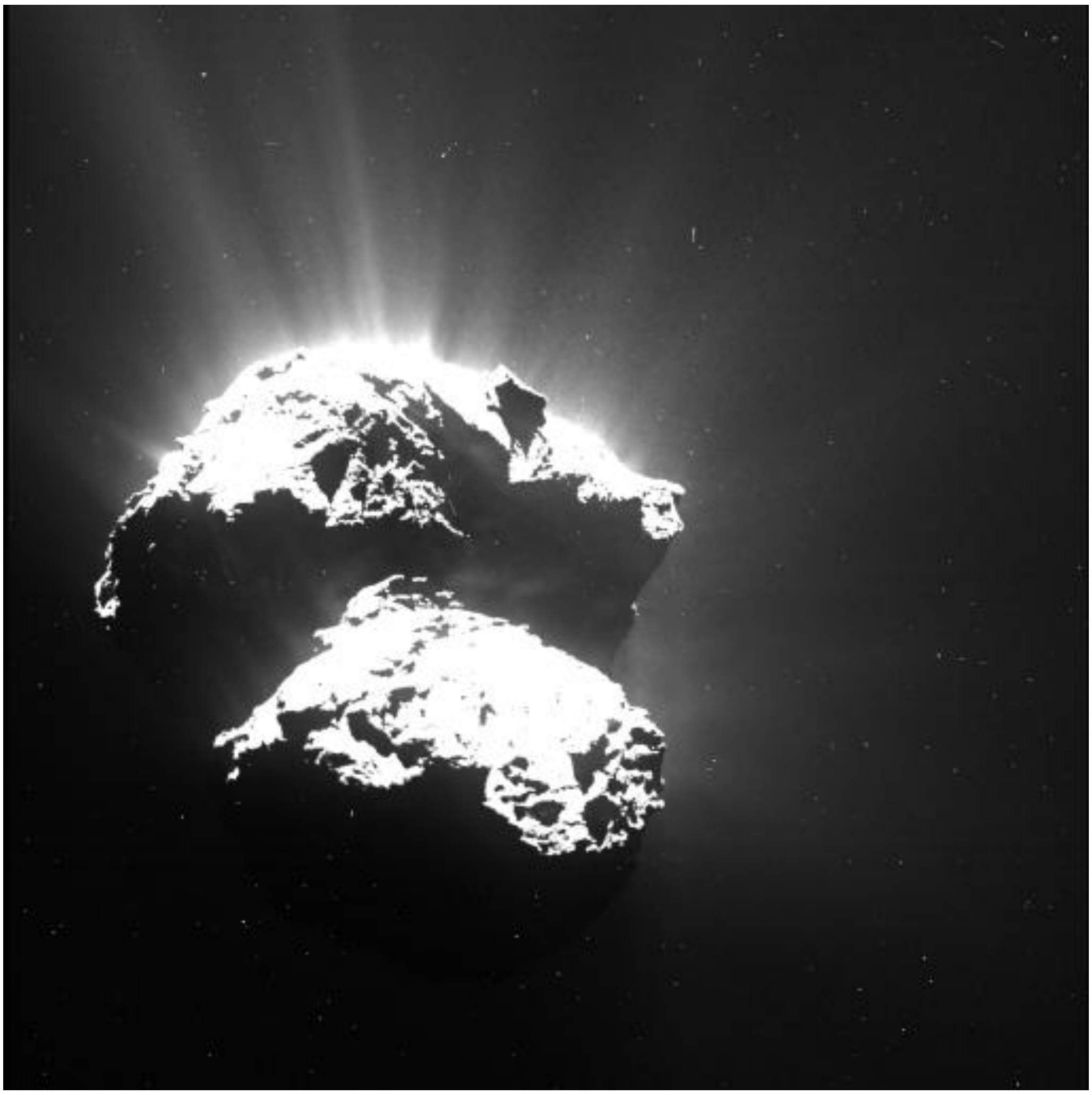

Figure 15. Image of Comet 67P/Churyumov–Gerasimenko captured by Rosetta's OSIRIS narrow-angle camera on 12 August 2015 (17:35 GMT) at a distance of 332 km from the comet nucleus. Dust emission and an outburst (cf. Vincent et al., 2016) is visible on the Sun-lit (top) side of the nucleus (ESA/Rosetta/MPS for OSIRIS Team MPS/UPD/LAM/IAA/SSO/INTA/ UPM/DASP/IDA).

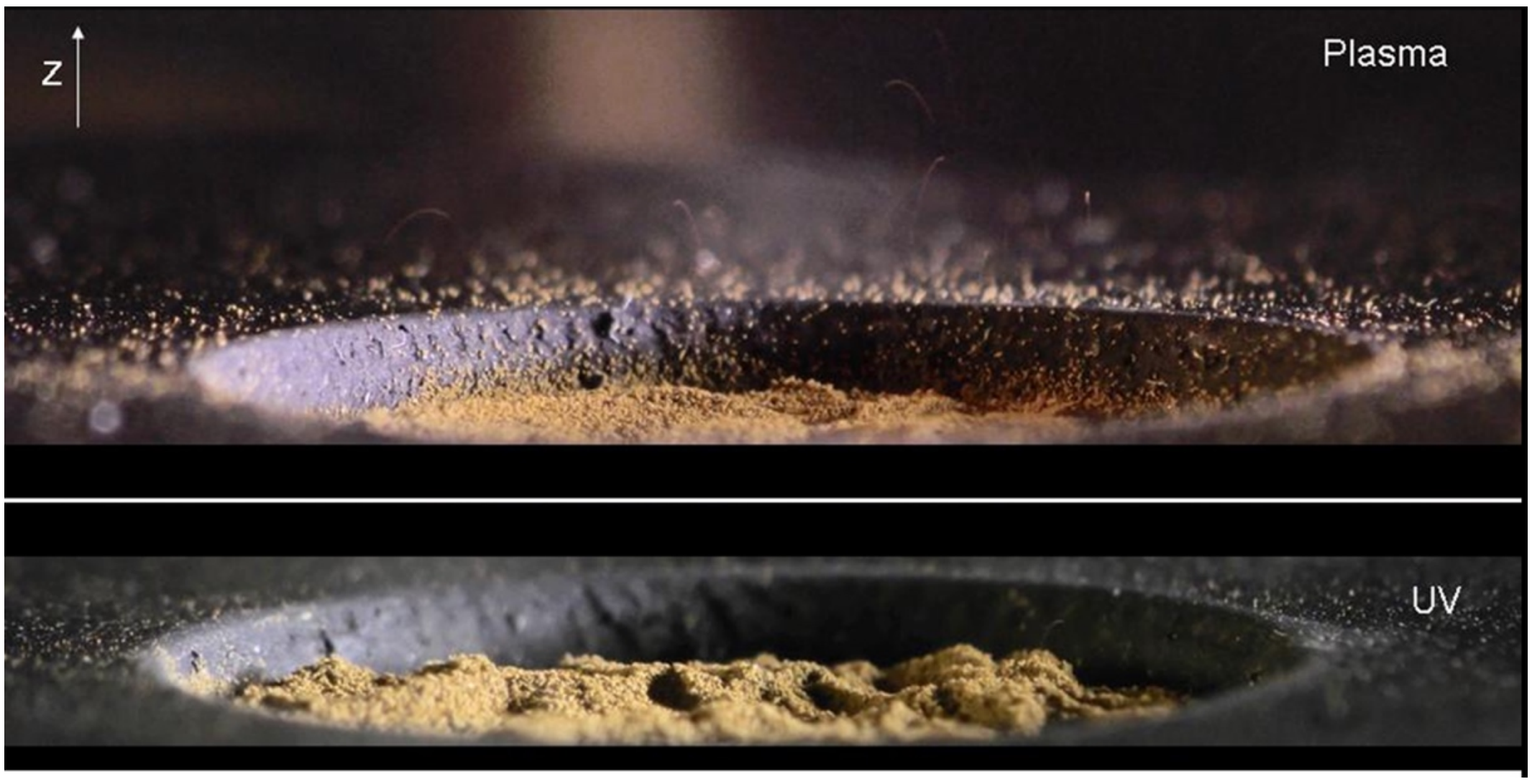

Figure 16. Dust hopping from a central reservoir during exposure to plasma and UV radiadtion (Wang et al. 2016).

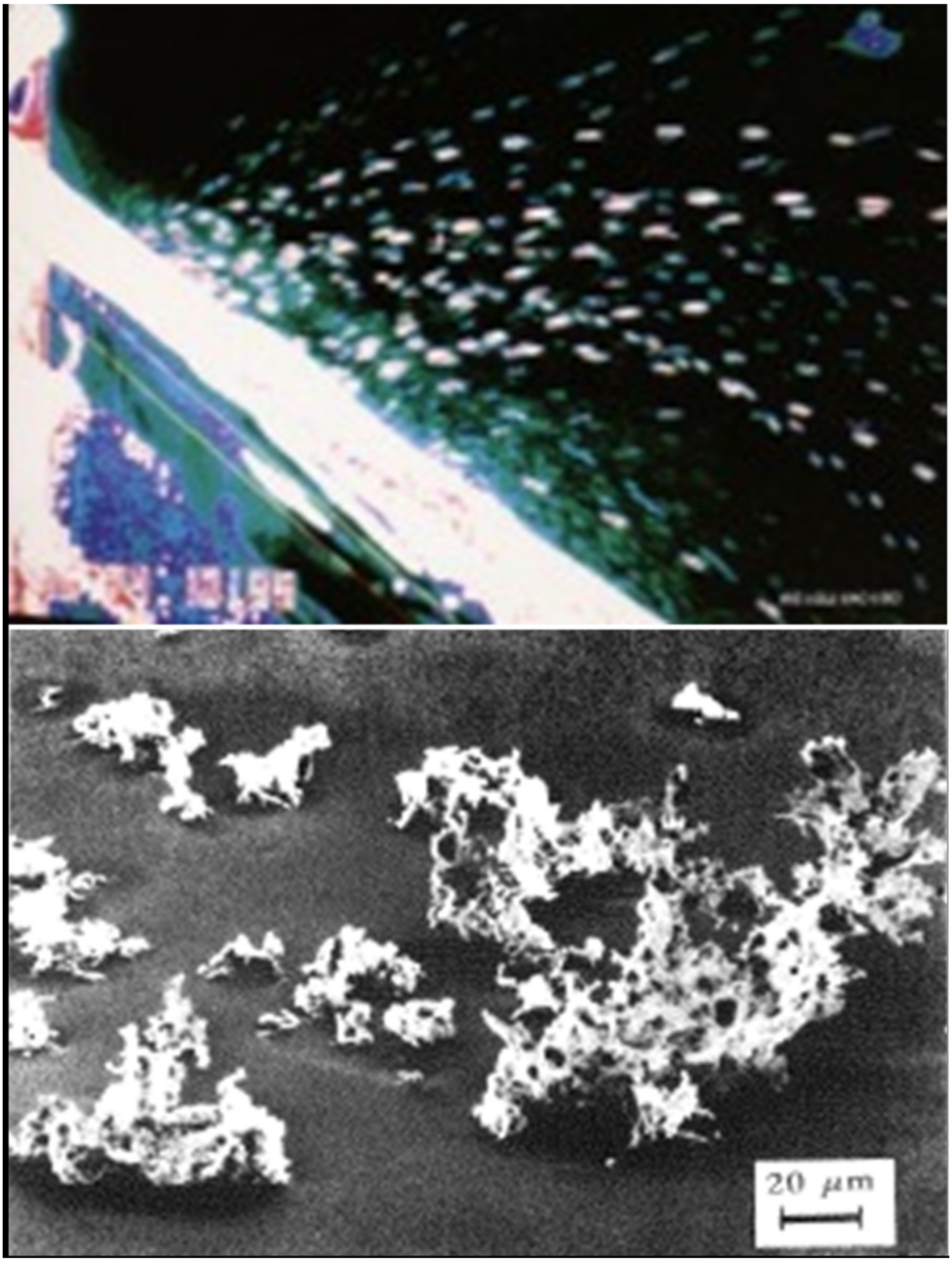

Figure 17. Dust release from an icy dust mixture during the KOSI Experiments (left) and collected dust particles (right, Kochan et al., 1998)

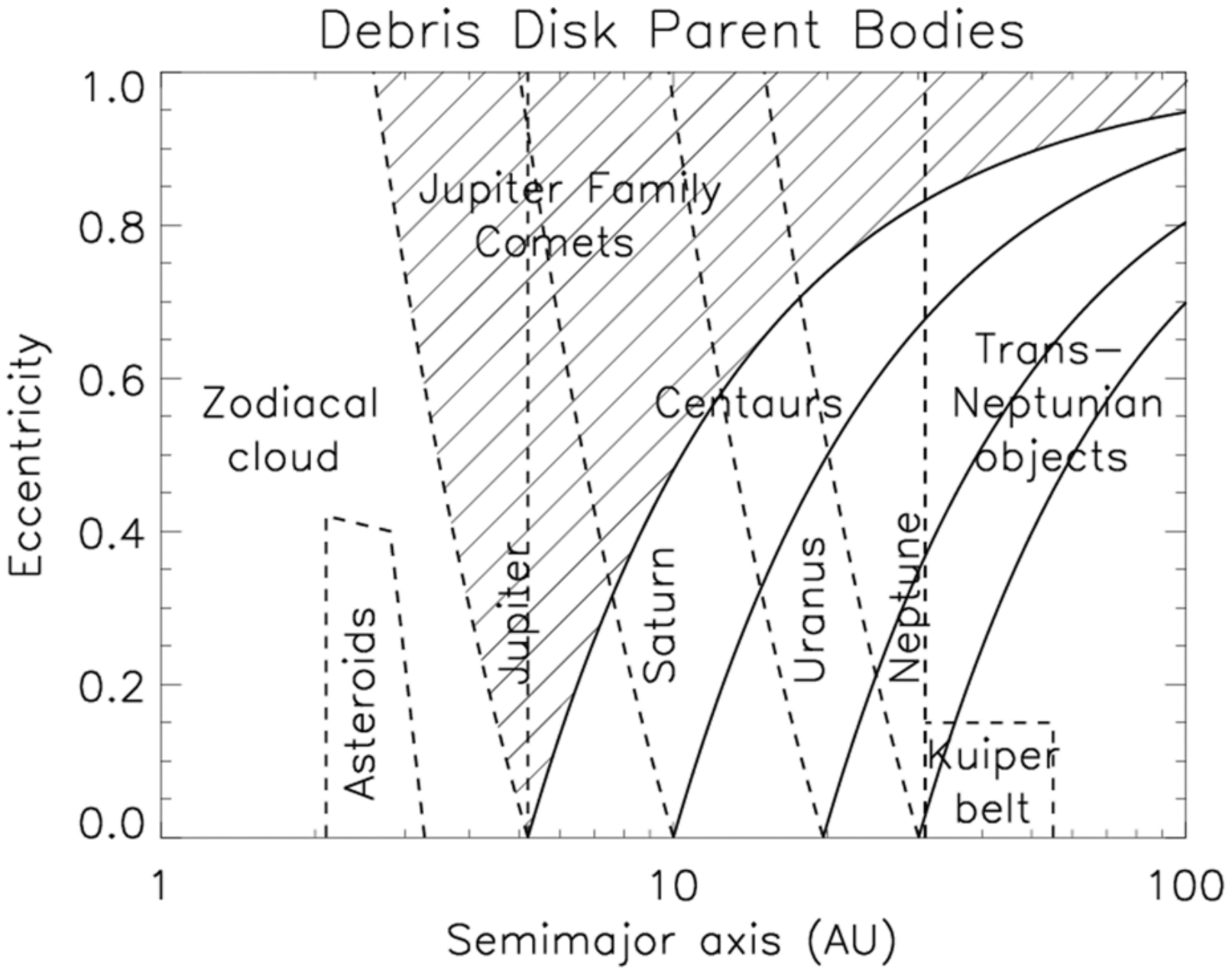

Figure 18. Relationship between small bodies in the solar system and the giant planets. The triangular-shaped regions pointing at the semi-major axes of the giant planets delineate the scattering zones of the respective planet.

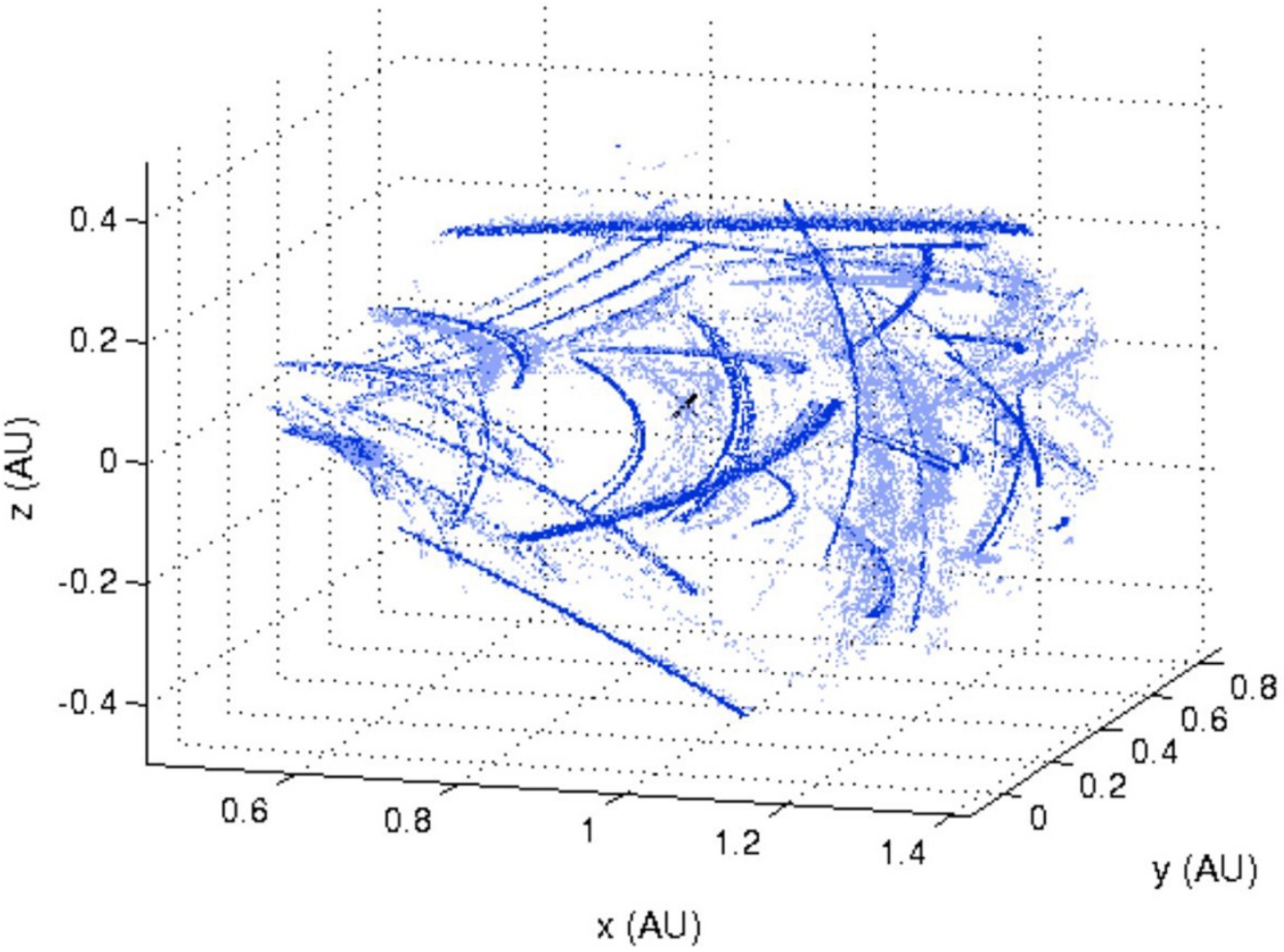

Figure 19. Prediction of the dust trail structure within 0.5 AU from the Earth on 1st October 2049 (ESA IMEX study, Soja et al., 2014). Dust positions are given in heliocentric ecliptic coordinates. Dark blue particles have mass 1 × 10−6 kg and light blue particles have mass 1 × 10−9 kg. Earth is the black cross.